# Assessment and Application of Wavelet-based Optical Flow Velocimetry (wOFV) to Wall-Bounded Turbulent Flows

Alexander Nicolas[*,1], Florian Zentgraf[2], Mark Linne[1], Andreas Dreizler[2], Brian Peterson[1]

[1] *The University of Edinburgh, School of Engineering, Institute of Multiscale Thermofluids, United Kingdom*
[2] *Technical University of Darmstadt, Department of Mechanical Engineering, Reactive Flows and Diagnostics, Germany*

**Abstract**

The performance of a wavelet-based optical flow velocimetry (wOFV) algorithm to extract high accuracy and high resolution velocity fields from tracer particle images in wall-bounded turbulent flows is assessed. wOFV is first evaluated using synthetic particle images generated from a channel flow DNS of a turbulent boundary layer. The sensitivity of wOFV to the regularization parameter ($\lambda$) is quantified and results are compared to cross-correlation-based PIV. Results on synthetic particle images indicated different sensitivity to under-regularization or over-regularization depending on which region of the boundary layer is being analyzed. Nonetheless, tests on synthetic data revealed that wOFV can modestly outperform PIV in vector accuracy across a broad $\lambda$ range. wOFV showed clear advantages over PIV in resolving the viscous sublayer and obtaining highly accurate estimates of the wall shear stress and thus normalizing boundary layer variables. wOFV was also applied to experimental data of a developing turbulent boundary layer. Overall, wOFV revealed good agreement with both PIV and a combined PIV + PTV method. However, wOFV was able to successfully resolve the wall shear stress and correctly normalize the boundary layer streamwise velocity to wall units where PIV and PIV + PTV showed larger deviations. Analysis of the turbulent velocity fluctuations revealed spurious results for PIV in close proximity to the wall, leading to significantly exaggerated and non-physical turbulence intensity in the viscous sublayer region. PIV + PTV showed only a minor improvement in this aspect. wOFV did not exhibit this same effect, revealing that it is more accurate in capturing small-scale turbulent motion in the vicinity of boundaries. The enhanced vector resolution of wOFV enabled improved estimation of instantaneous derivative quantities and intricate flow structure both closer to the wall and more accurately than the other velocimetry methods. These aspects show that, within a reasonable $\lambda$ range that can be verified using physical principles, wOFV can provide improvements in diagnostics capability in resolving turbulent motion occurring in the vicinity of physical boundaries.

## 1. Introduction

Fluid flow dynamics and the interaction with walls are of prime importance in a variety of engineering applications. The dynamics of the 'boundary layer' region are of major interest and have been the subject of extensive research since the fundamental work of (Prandtl, 1904). Detailed knowledge of the momentum transport processes within turbulent boundary layers underpins the success of many industrial, aerodynamic and medical designs and their relevant applications. Obtaining accurate velocity measurements across the extent of the boundary layer flow is key to developing a sound understanding of the complex multiscale phenomena present in wall-bounded turbulence. The structure of the turbulent boundary layer is commonly delineated based on regions primarily dominated by either viscous stresses (the viscous sublayer), turbulent Reynolds stresses (the logarithmic region) or influenced by both (the buffer layer). To evaluate and facilitate comparison of these different regions between theoretical, numerical and experimental results, the boundary layer mean streamwise velocity $\langle U_1 \rangle$ and wall-normal distance coordinate $x_2$ are typically normalised to so-called "wall units":

$$u^+ = \frac{\langle U_1 \rangle}{u_\tau} \qquad (1)$$

$$y^+ = x_2 \frac{u_\tau}{\nu} \qquad (2)$$

where $\nu$ is kinematic viscosity, a physical property of the fluid. The key variable involved in the nondimensionalization is the friction velocity $u_\tau$, defined as:

---

[*] Corresponding author: alexander.nicolas@ed.ac.uk

$$u_\tau = \sqrt{\frac{\tau_w}{\rho}} \qquad (3)$$

where $\rho$ is the fluid density. $\tau_w$ is the mean wall shear stress:

$$\tau_w = \mu \left.\frac{\partial \langle U_1 \rangle}{\partial x_2}\right|_{x_2=0} = \mu\gamma \qquad (4)$$

where $\mu$ is the dynamic viscosity ($\mu = \nu\rho$). For flows with constant physical properties, it can be seen from Eqns. 1-4 that the normalising variable $u_\tau$ is ultimately defined by the wall shear stress $\tau_w$ and therefore by the estimate of the velocity gradient at the wall $\gamma$. The accurate determination of $\gamma$ is thus crucial for accurate scaling and subsequent evaluation of boundary layer quantities. This sharp gradient due to the no-slip condition at the wall is challenging to resolve experimentally due to the need to sample flow motion down to the wall with sufficient accuracy with minimal disturbance to the flow itself.

Non-intrusive flow measurement techniques such as digital particle image velocimetry (PIV) have become well established for boundary layer investigations (Adrian et al., 2000; Willert, 2015; De Silva et al., 2014; Dennis & Nickels, 2011; Gao et al., 2013; Herpin et al., 2008; Lehew et al., 2013; Schröder et al., 2011). To determine the velocity, each PIV image is subdivided into interrogation windows (IWs), which are cross-correlated between image frames. PIV has become a mature diagnostic technique that is robust, efficient and well-understood in terms of its sources of error and theoretical underpinnings. However, a fundamental limitation still exists in that the spatial resolution of PIV is directly related to the smallest size of the IW (Kähler et al., 2012a). Since the velocity vector represents a spatially-averaged velocity of particles within each IW, the estimated velocity is a low-pass filtered version of the true fluid flow, which is problematic if turbulent fluctuations and velocity gradients are present within the IW itself. Particularly in the case of wall-bounded flows, which always feature strong velocity gradients near the wall, obtaining accurate and reliable velocity measurements in the vicinity of the wall can present challenges for cross-correlation-based PIV. The low-pass filtering effect increases uncertainty in regions of high velocity gradients due to an increased spread and biasing of the correlation peak (Scarano & Riethmuller, 2000; Kähler et al., 2012; Raffel et al., 2018).

Another velocimetry technique is particle tracking velocimetry (PTV) which attempts to detect and subsequently match individual tracer particles between frames to determine their velocity. This method is sometimes used as a subsequent step following an initial PIV result in hybrid PIV + PTV algorithms (Keane et al., 1995; Stitou & Riethmuller, 2001). Use of PIV + PTV can significantly improve the achievable spatial resolution over PIV (Kähler et al., 2012a), without requiring low seeding densities and has been employed to study boundary layer flows (Renaud et al., 2018; Ding et al., 2019; Kähler et al., 2012b). However, PTV vector fields often contain higher noise levels in the signal, and sufficient filtering or direct spatial/ensemble averaging is required to mitigate this noise.

A promising alternative to these traditional velocimetry techniques is a method originating from the field of computer vision, known as optical flow (Horn & Schunck, 1981). Optical flow is a method often referred to in literature as dense motion estimation, i.e., a velocity vector is calculated for every pixel in a digital image. Application of optical flow velocimetry (OFV) techniques have previously demonstrated increased accuracy and resolution over conventional correlation-based methods (Yuan et al., 2007; Ruhnau et al., 2007; Corpetti et al., 2006; Héas et al., 2012; Dérian et al., 2013; Kadri-Harouna et al., 2013; Schmidt & Sutton, 2020; B. E. Schmidt et al., 2021). Such studies have primarily focused on synthetic and experimental test cases involving analytical flows, isotropic turbulence and free shear flows.

The impressive spatial resolution and improved velocity vector accuracy associated with OFV makes it an attractive tool to resolve velocities close to the wall, enabling reliable calculation of $\gamma$. At the same time, OFV can improve estimation of small-scale turbulent fluctuations near the wall as well as computation of derivative quantities which yield insight on near-wall vortical structures that are believed to play an important role in the organization of turbulence within the boundary layer (Robinson, 1991; Herpin et al., 2012; Adrian et al., 2000). Despite its capabilities, only a few applications of OFV to wall-bounded flows exist in the literature (Kapulla et al., 2011; Kähler et al., 2016; Stanislas et al., 2005; Ruhnau & Schnörr, 2006; Stark, 2013; Cai et al., 2019; Gevelber et al., 2022). Such studies primarily use wall-bounded environments as test cases for other aspects of the specific OFV algorithms and limit investigations to velocity profiles. A thorough evaluation of OFV to resolve a turbulent boundary layer is limited. Furthermore, analysis of derived quantities such as the wall shear stress as well as evaluation of the accuracy and effect on resolution of the inner-scaled turbulent boundary layer quantities is absent in the literature.

Variational OFV techniques involve selection of a scalar regularization parameter that is typically determined empirically. Regularization imparts a degree of spatial regularity to the estimated flow field that suppresses non-physical noise and provides closure to the optical flow problem. Correct selection of the regularization parameter $\lambda$ is key in obtaining accurate velocity fields that accurately resolve fine-scale motion without excessive damping or smoothing of velocity gradients. This is especially important in estimating $\gamma$, where the discontinuity in motion at the wall can be particularly susceptible to the smoothing effect inherent in regularization (Weickert & Schnörr, 2001; Kalmoun, 2018; Zach et al., 2007; Black & Anandan, 1996; Aubert, 1999). To the best of the authors' knowledge, other works exploring or discussing this parameter in the context of fluid velocimetry are limited to those of (Corpetti et al., 2002; Kapulla et al., 2011; Stark, 2013; Schmidt & Sutton, 2020; Cai et al., 2018; Heás et al., 2013). None of these, however, have investigated the sensitivity of $\lambda$ specifically in relation to near-wall measurements in wall-bounded flows.

The present work assesses the performance of an advanced wavelet-based optical flow velocimetry (wOFV) method to obtain highly resolved and accurate measurements of velocity and derived quantities such as the wall shear stress in turbulent wall-bounded flows. The influence of regularization on velocity results and normalized boundary layer quantities is investigated to understand the effect of this parameter. The first part of the manuscript provides an overview of optical flow and a brief outline of the wavelet-based implementation. This is followed by a detailed assessment and sensitivity study of the regularization parameter on wOFV results in comparison to correlation-based PIV using synthetic particle images generated from DNS of a turbulent channel flow. The final part of this work applies wOFV to an experimental PIV dataset featuring a developing turbulent boundary layer. Results are compared to correlation-based PIV processing to demonstrate the advantages of wOFV as an alternative technique in the study of turbulent wall-bounded flows.

**2. Optical Flow**
**2.1 Principles**

Optical flow describes the apparent displacement of brightness intensity patterns in an image sequence (Horn & Schunck, 1981). The basic assumption in optical flow techniques is the conservation of a quantity in the image plane, typically brightness intensity along a point trajectory. This is expressed as the optical flow constraint equation (OFCE):

$$\frac{dI(\pmb{x},t)}{dt} = \frac{\partial I(\pmb{x},t)}{\partial t} + \pmb{U}(\pmb{x},t) \cdot \nabla I(\pmb{x},t) = 0 \qquad (5)$$

where $I(\pmb{x},t)$ is the brightness intensity at pixel locations $\pmb{x} = (x_1, x_2)^T$ in the image domain $\Omega$ and $\pmb{U}(\pmb{x},t) = \left(U_1(\pmb{x},t), U_2(\pmb{x},t)\right)^T$ is the two-dimensional displacement. Equation 5 is recognisable as a transport equation of a passive scalar in a divergence-free flow (Liu & Shen, 2008). Assuming a constant velocity and a unit time interval between the image pair, Eq. 5 can be integrated to the displaced frame difference (DFD):

$$I_0(\pmb{x}) - I_1(\pmb{x} + \pmb{U}(\pmb{x})) = 0 \qquad (6)$$

Equation 5 or 6 is known as the *data term* in OF literature. It establishes the relationship between a measurement in the image plane $I(\pmb{x},t)$ and the variable to be calculated $\pmb{U}(\pmb{x})$. The data term is incorporated into a penalty function to be minimised, commonly a quadratic penalty as employed in the present study:

$$J_D = \int_\Omega [I_0(\pmb{x}) - I_1(\pmb{x} + \pmb{U}(\pmb{x}))]^2 \, d\Omega \qquad (7)$$

The data term however is ill-posed, as it relates a two-dimensional velocity to only one observed variable being the image intensity. This results in an ambiguous situation where only motion perpendicular to brightness gradient contours can be determined, known in literature as the aperture problem (Beauchemin & Barron, 1995). Different methods of resolving the aperture problem exist (Barron et al., 1994). In the seminal work of (Horn & Schunck, 1981), a variational approach was proposed to assimilate the data term together with an additional smoothness constraint known as the *regularization term* $J_R$, weighted by a scalar parameter $\lambda$, into a minimization problem to solve for the image plane per-pixel displacement:

$$\widehat{\pmb{U}} = \arg\min_{\pmb{U}} J_D(I_0, I_1, \pmb{U}) + \lambda J_R(\pmb{U}) \qquad (8)$$

where the caret ( ˆ ) denotes the final estimated quantity. The regularization term, which is solely a function of the velocity field, provides additional information about the velocity field to compensate for regions where motion cannot be determined from the data term alone, such as tangentially along image contours, as well as regions of constant, uniform image features where spatiotemporal image gradients vanish. $J_R$ affects the spatial coherence of neighbouring velocity vectors and enforces a degree of regularity or visually perceived "smoothness" to the velocity field. Regularization also functions as a type of outlier rejection process during the minimization (Heitz et al., 2010) and reduces the susceptibility of the estimated vector field to noise and imaging imperfections. In the context of fluid velocimetry, regularization terms involving higher-order derivatives of the velocity field are preferred to better preserve velocity gradients and thus improve estimation of derived quantities such as vorticity and strain-rate. The present work uses the Laplacian regularization in a quadratic penalty:

$$J_R = \int_\Omega |\nabla^2 U_1|^2 + |\nabla^2 U_2|^2 \, d\Omega \qquad (9)$$

with the continuous wavelet operator approximation described in (Kadri-Harouna et al., 2013). Laplacian regularization imparts a physically sound smoothing in a similar manner to viscosity in divergence-free two-dimensional flows (Schmidt & Sutton, 2021). Using the Laplacian regularization provides nearly identical accuracy but with significantly less computing time than other high-order schemes such as the second-order divergence curl (Suter, 1994) or viscosity-based regularization (Schmidt & Sutton, 2021).

Variational optical flow techniques seek a per-pixel vector field transformation $\widehat{U}$ that maps one image onto the subsequent that best: (1) conserves pixel brightness intensity and (2) enforces the regularity defined by $J_R$. The parameter $\lambda$ in Eq. 8 establishes the relative importance of $J_D$ versus $J_R$ during the minimization process and determines the extent to which $J_R$ can deviate the estimated velocity field $\widehat{U}$ from the constraint of brightness conservation in $J_D$. Lower $\lambda$ values place a stronger emphasis on reducing $J_D$ during minimization, thus attempting to better match pixel intensities between $I_0$ and $I_1$ to reduce the DFD even if the intensity variations do not correspond to the true motion. This creates non-physical velocity fluctuations at fine scales visible as noise in $\widehat{U}$. Increasing $\lambda$ dampens the small-scale motion, however, a higher $\lambda$ weighting can lead to excessive smoothing of the velocity field. Sensitivity analysis of this parameter is a key aspect in understanding the applicability of OFV as an alternative diagnostic technique to studying wall-bounded flows.

## 2.2 Wavelet-based Optical Flow

The current wOFV implementation was proposed by Dérian (2013), developing on the original wavelet-based optical flow methods of Wu et al. (2000) and Chen et al. (2002) for computer vision applications. Improvements in the form of symmetric boundary conditions (Schmidt & Sutton, 2020) and efficient implementation of high-order and physically-sound regularization terms (Kadri-Harouna et al., 2013; Schmidt & Sutton, 2021) have furthered the robustness and accuracy of this technique. For brevity, only an overview of the wavelet-based optical flow method is presented. Details of the wOFV algorithm can be found in (Dérian et al., 2013; Kadri-Harouna et al., 2013; Schmidt & Sutton, 2019; Schmidt & Sutton, 2020).

The principle of wOFV, in contrast to other OFV techniques, is to perform the minimization in Eq. 8, not over the physical velocity field $U(x)$, but over the wavelet coefficients $\boldsymbol{\theta} = (\theta_1, \theta_2)^T$ from its Discrete Wavelet Transform (DWT) $\boldsymbol{\theta} = \boldsymbol{\Psi}^{-1}(x)U(x)$, where $\boldsymbol{\Psi}^{-1}(x)$ denotes the wavelet transform decomposition operator. The minimization problem is then expressed as:

$$\widehat{\boldsymbol{\theta}} = \arg \min_{\boldsymbol{\theta}} J_D(I_0, I_1, \boldsymbol{\theta}) + \lambda J_R(\boldsymbol{\theta}) \qquad (10)$$

Broadly speaking, a wavelet transform extracts the frequency content of a signal (or image in 2D) at different scales of resolution (Mallat, 2009). The wavelet transformed velocity field coefficients $\boldsymbol{\theta}$ are optimized sequentially in a multi-resolution strategy. The wavelet coarsest-scale coefficients are estimated first, before estimating coefficients associated with progressively finer scales until the pixel scale is reached. Previous coarse-scale velocity estimates are included in every level of estimation, therefore earlier spurious vectors from coarser-scale velocity estimates are corrected for as finer-scale motion is determined. Once the full minimization is complete, the velocity field in physical space is recovered by application of the DWT reconstruction operator $\boldsymbol{\Psi}(x)$ to the output wavelet coefficients $\boldsymbol{\Psi}(x)\widehat{\boldsymbol{\theta}} = \widehat{U}$.

To cope with large displacements, traditional OFV methods commonly use multiresolution coarse-to-fine warping strategies (Heitz et al., 2010), which extend the achievable dynamic range. This approach, however, can lead to propagation of errors during the multi-scale estimation process with no possibility of posterior correction. Conversely, the multiresolution framework inherent in wavelet decompositions provides a natural scheme that is well-suited to represent the multi-scale nature of turbulence (Deriaz & Perrier, 2009; Farge et al., 1996). Decomposition of the velocity field across the wavelet basis functions also allows for accurate implementation of high-order derivatives (Beylkin, 1992) used in the calculation of $J_R$. Previous studies demonstrated wOFV to be among some of the more accurate existing modern OFV methods, see (Cai et al., 2018; Kadri-Harouna et al., 2013; Dérian et al., 2013; Schmidt & Sutton, 2019).

In this work, the wOFV implementation uses the odd length biorthogonal nearly-coiflet wavelet basis (BNC 17/11) introduced by (Winger & Venetsanopoulos, 2001). This basis has a *nearly* maximum number of vanishing moments possible for a given biorthogonal wavelet filter size. Maximising the number of vanishing moments increases velocity estimation accuracy up to a degree (Dérian et al., 2013). This wavelet family is notable for the improved retention of fine details in its wavelet transform partial reconstructions which are implicit in the multi-scale estimation process in wOFV methods. Since the basis is biorthogonal, it is implemented using the non-expansive symmetric boundary condition described in (Schmidt & Sutton, 2020) which eliminates boundary artefacts resulting from a lack of periodicity in the imaged motion.

## 3. Description of Synthetic Test Case

In order to quantitatively assess wOFV performance and $\lambda$ parameter sensitivity in wall-bounded flows, it is first necessary to compare estimated velocity fields from wOFV to a known ground truth velocity available from synthetic data. Synthetic data provides a useful test platform where parameters can be easily and independently controlled in an idealized image environment. In this work, the synthetic data are derived from direct numerical simulation (DNS) of a turbulent channel flow (Graham et al., 2016) hosted online at John Hopkins Turbulence Database (JHTDB) (Li et al., 2008). Details of the synthetic data are described below.

### 3.1 DNS dataset

Table 1 describes the simulation parameters of the JHTDB, which are stored in a nondimensional form based on the half-channel height $h$. 100 temporally correlated velocity fields are extracted with a time separation of 2.5 $\delta t$ (stored) DNS database timesteps from a subset of the DNS domain that includes the no-slip velocity grid point of the lower wall. The fields are of nondimensional size $0.17h \times 0.17h \times (5 \times 10^{-4})h$ and sampled from the database at a grid resolution of $1024 \times 1024 \times 3$ using fourth-order Lagrange polynomial interpolation (Berrut & Trefethen, 2004).

| | |
|---|---|
| Bulk velocity $U_b$ | 0.99994 |
| Centreline velocity $U_c$ | 1.1312 |
| Friction velocity $u_\tau$ | 0.0499 |
| Kinematic viscosity $\nu$ | $5 \times 10^{-5}$ |
| Bulk velocity Reynolds number $Re_b = U_b 2h/\nu$ | $3.9998 \times 10^4$ |
| Centreline velocity Reynolds number $Re_c = U_c h/\nu$ | $2.2625 \times 10^4$ |
| Friction velocity Reynolds number $Re_\tau = u_\tau h/\nu$ | 999.35 |
| DNS database timestep $\delta t$ | 0.0065 |
| Full domain size | $8\pi h \times 2h \times 3\pi h$ |
| Full grid resolution | $2048 \times 512 \times 1536$ |

Table 1: Simulation parameters (nondimensional) of the JHTDB channel flow DNS.

### 3.2 Particle Image Generation

Once the velocity fields from the DNS are extracted, it is necessary to determine tracer particle displacements between frames of each image pair as they are advected by the DNS velocity. For the initial frame of each image pair, synthetic particle tracer locations are initialized from a random distribution for each of the extracted DNS velocity fields. The velocity field is assumed to be constant between consecutive images and the tracers assumed to be spherical and massless. The displacement of each particle in each second frame is computed numerically using an explicit Runge-Kutta scheme (Dormand & Prince, 1980) and a modified Akima spline interpolation for the velocities at particle locations (Akima, 1974). The velocity fields and particle displacements are then scaled from the nondimensional DNS units to pixel displacements per unity interframe time interval ($dt = 1$) such that the maximum image plane velocity magnitude corresponds to ~3.5 $px/dt$, and the maximum out-of-plane displacement is ~0.8 $px/dt$.

The particle image pixel intensities are determined using classical methods of synthetic particle image generation (Raffel et al., 2018). The maximum particle intensity is governed by its diameter $d_p$ and out-of-plane position $x_{3,p}$ within a Gaussian profile synthetic laser sheet:

$$I_p = d_p^2 \exp\left(\frac{-8(x_{3,LS} - x_{3,p})^2}{2\sigma_{LS}^2}\right) \tag{11}$$

The laser sheet position $x_{3,LS}$ is centred in the middle of the extracted DNS domain. The standard deviation of the laser sheet profile is set to $\sigma_{LS} = 2$ such that the $1/e^2$ profile thickness is equal to the out-of-plane $x_3$ thickness of the DNS volume subsection. In this way, the out-of-plane particle displacement is less than 1/4 of the laser sheet thickness as recommended by (Adrian & Westerweel, 2011). Each particle is randomly assigned a diameter that is drawn from a log-normal distribution of values:

$$PDF = \frac{1}{x\sigma\sqrt{2\pi}} exp\left(\frac{-(\log x - \mu)^2}{2\sigma^2}\right) \tag{12}$$

with parameters $\mu = 0.90\ px$ and $\sigma = 0.76\ px$ for the mean and standard deviation, respectively. The particle seeding density is 0.03 particles per pixel$^2$ (PPP), representative of that estimated from the experimental data presented in Sect. 5. The in-plane pixel intensity is computed from the integral form of the Gaussian function solved analytically using error functions. This is a more representative method of how a camera integrates the light intensity over individual pixels compared to simply using the analytical Gaussian expression. Finally, after the pixel intensities have been determined, the values are scaled to the dynamic range of an 8-bit camera sensor and rounded to integers to mimic discretisation.

Once the particle images are rendered, the images are vertically shifted upwards by 160 pixels to create a masked wall region of zero intensity. This vertical shift avoids having the flow region near the bottom of the image where boundary conditions in the wavelet transforms of wOFV can affect the near-wall velocity estimates. Moreover, this shift of the flow region from the image boundary is consistent with experimental images presented in Sect. 5. As this masked region in the images has 0 intensity, it does not contribute to the DFD and is effectively ignored in the minimization (Schmidt & Woike, 2021). An example of a rendered particle field image is shown in Fig. 1.

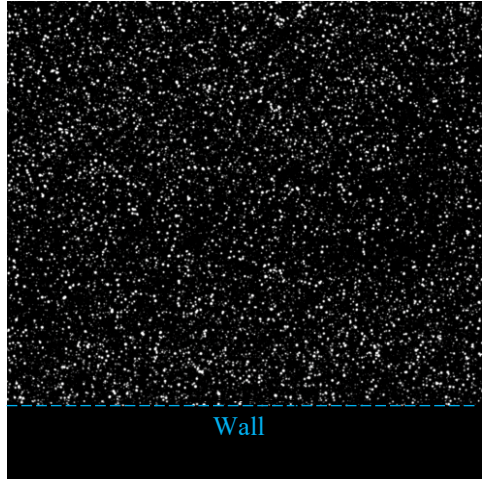

**Fig. 1** Example rendered particle field image from the channel flow DNS.

## 4. wOFV assessment using synthetic data

In this section, wOFV performance is assessed using the ground truth DNS data for comparison. In particular, the sensitivity of the wOFV results to the regularization weighting $\lambda$ is evaluated to understand how $\lambda$ selection affects estimation of turbulent boundary layer motion. wOFV findings are reported for 6 values of $\lambda = [2, 40, 100, 180, 520, 1000]$. This range of $\lambda$ covers velocity estimates ranging from under-regularized (visibly noise-dominated) to over-regularized (over-smoothed). The wOFV error throughout the $\lambda$ range is determined to identify suitable $\lambda$ values where the wOFV error outperforms PIV. In Sect. 4.1, characteristic velocity estimates for each $\lambda$ are presented together with the error over the entire image domain. Sect. 4.2 evaluates the effect of $\lambda$ on the calculation of wall units and the effect this has on the interpretation of the mean velocity within each region of the boundary layer.

PIV is also applied to the synthetic data, providing a benchmark to compare wOFV with the current state-of-the-art. A commercial cross-correlation-based PIV software (DaVis 10.0.05, LaVision) was used for PIV processing. The cross-correlation algorithm used 2 and 3 passes for the initial and final Gaussian-weighted IWs of size 64 x 64 down to 16 x 16 with 75% overlap. The anisotropic denoising filter in DaVis was applied to the PIV vector fields. The filter strength was selected for the most accurate results for the given IW size. Thus, it should be strongly emphasized that the PIV results presented are *optimized*. A geometric mask was placed 1 pixel below the no-slip grid point to capture the entire particle image region. This results in the first PIV velocity vector being 11 pixels above the no-slip pixel ($x_2 = 0$). For both PIV and wOFV the particle images were preprocessed using a min-max filter (Adrian & Westerweel, 2011) to account for changes in particle intensity resulting from out-of-plane motion within the synthetic laser sheet.

A notable feature of wOFV is its ability to provide dense velocity estimates with per-pixel vector spacing. Although this impressive vector spacing is achievable, the true spatial resolution of wOFV is a subject not often discussed and requires thorough analysis which is beyond the scope of this work. The average spacing between particles can be considered to be a conservative estimate of wOFV spatial resolution, since this is the average maximum distance between image features containing a genuine intensity signal. In the synthetic data the average spacing between particles is 5.8 pixels. This value is considered as an upper bound for wOFV's spatial resolution, as this estimate only considers the average distance between particle centres and does not take into account each particles' local intensity distribution for which additional valid vectors are associated. Additionally, because an explicit regularization scheme is used, the vectors in regions without particles will contain physically-sound flow field information from regions containing genuine signals (Schmidt & Sutton, 2021). Such features would decrease the true spatial resolution, but this requires further analysis. Thus, we report 5.8 pixels as the spatial resolution for wOFV, while vectors are resolved per pixel. The PIV spatial resolution is reported as the final IW size (i.e., 16 pixels), while the vector spacing is 4 pixels.

### 4.1 $\lambda$ sensitivity based on entire image domain
#### 4.1.1 Single image analysis

The influence of $\lambda$ is first described by evaluating features of the wOFV velocity field within the entire image domain. Vector accuracy over the entire image domain is assessed by the normalized root mean square error:

$$\varepsilon_u = \sqrt{\frac{1}{n_v}\sum \frac{(U_1 - U_{1,DNS})^2 + (U_2 - U_{2,DNS})^2}{U_{1,DNS}^2 + U_{2,DNS}^2}} \qquad (13)$$

In Eq. 13, $n_v$ is the total number of vectors and $U_i$ is the individual velocity value in the streamwise and normal direction denoted by subscripts $i = 1, 2$ respectively. Normalization by the DNS velocity magnitude ensures that errors in regions of very low velocities near the wall are properly accounted for and not dominated by errors from large velocity magnitudes (McCane et al., 2001). Vectors from wOFV outside the PIV masked boundary are ignored in the error calculation of wOFV for equivalent comparison. For comparison with PIV, the DNS ground truth velocity is subsampled to a lower resolution grid using spline interpolation.

Figure 2 shows the instantaneous velocity field magnitude from a subset of 4 selected $\lambda$ values. For comparison, the true velocity field from DNS and the corresponding velocity field from PIV are also shown. The associated $\varepsilon_u$ value for each result is reported above each sub-figure. For wOFV with $\lambda = 2$, the velocity estimate is under-regularized, leading to fine-scale noise visible as a speckle-like pattern within the velocity field and yields the highest $\varepsilon_u$ of the results shown. As $\lambda$ increases to 40, the regularity of the estimated flow field is increased and the noise becomes noticeably suppressed. At $\lambda = 180$, the noise is effectively removed from the velocity field and achieves the lowest $\varepsilon_u$, thus providing the closest agreement with the DNS. This $\lambda$ value producing the minimum $\varepsilon_u$ will be referred to as $\lambda^*$. Far beyond $\lambda^*$ at $\lambda = 1000$, the flow field is considered over-regularized; the noise has been eliminated entirely at the expense of over-smoothing the flow and therefore deviating from the DNS with $\varepsilon_u$ nearly doubling. PIV produces a high-quality velocity estimate with $\varepsilon_u$ as low as the $\lambda = 40$ wOFV result. With $\lambda^*$, a modest improvement of ~23% in $\varepsilon_u$ is achieved over PIV, demonstrating wOFV's improved accuracy over the state-of-the-art.

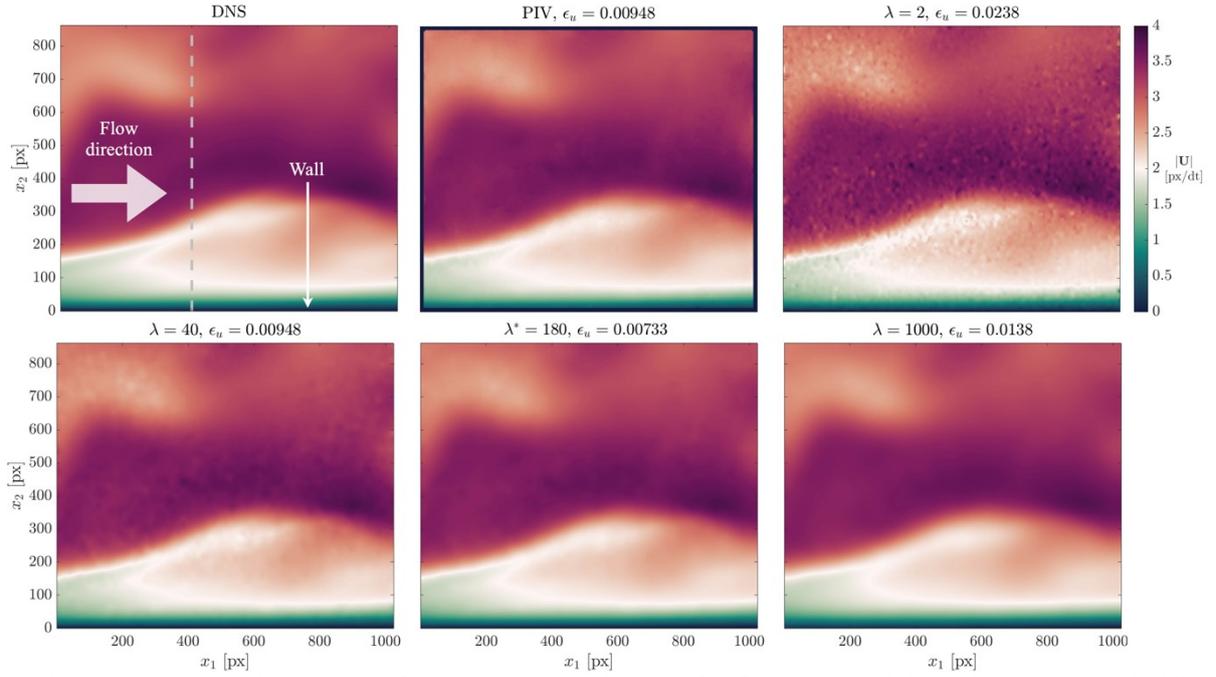

**Fig. 2** Instantaneous velocity magnitude for the DNS, PIV and wOFV ($\lambda = [2, 40, 180, 1000]$) results. The gray dashed line marks the location of the velocity profiles shown in Fig. 3.

Further assessment of how $\lambda$ influences the estimated wOFV velocity field is shown by evaluating local velocity profiles. Figure 3 shows the instantaneous streamwise velocity $U_1$ profiles extracted normal to the wall at pixel location $x_1 = 400$ marked by the gray dashed line in Fig. 2. This $x_1$ location was chosen arbitrarily but reveals trends consistent across all $x_1$ locations. The characteristic noise present for the under-regularized $\lambda = 2$ is clearly seen as spurious small-scale velocity oscillations. These fluctuations are reduced significantly as $\lambda$ is increased to $\lambda^*$, leading to overall better agreement with the DNS. In regions that contain high velocity gradients as shown near the inflection point at $x_2 = 240$ in Fig. 3b, it is shown that $\lambda = 40$ follows the DNS better than $\lambda^* = 180$. Thus, even though $\lambda^*$ is on average optimal for the entire imaged motion, localized regions of sharp velocity gradients may prefer a slightly lower $\lambda$ to avoid washing out small-scale flow features. This aspect is further discussed in Sect 4.1.3. As $\lambda$ exceeds $\lambda^*$, the over-smoothing effect is seen as a deviation from the DNS with velocity gradients becoming increasingly underestimated as clearly visible in Fig. 3b. As a benchmark, PIV processing achieves good agreement with the DNS, but with the reduced vector spacing (1 vector per 4 pixels) as visible in Fig. 3b.

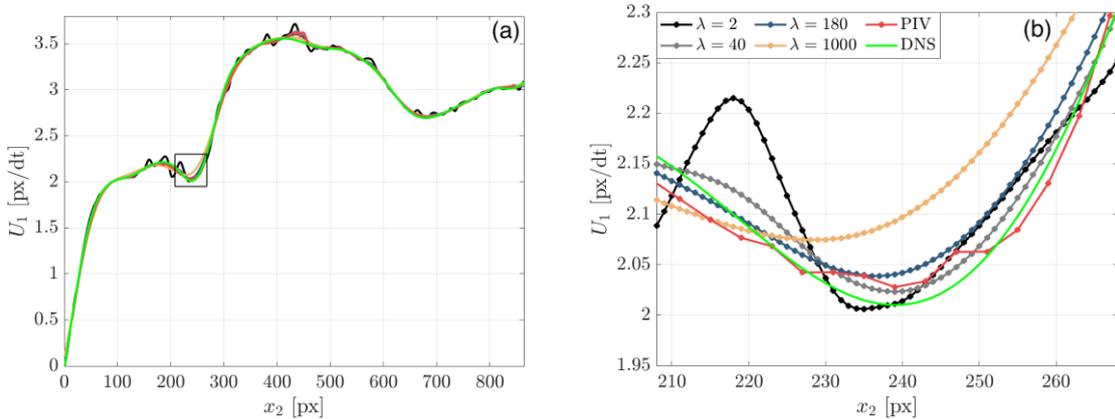

**Fig. 3 a** Velocity profile along grey line location in Fig. 2. **b** Zoomed view of high velocity gradient region within velocity profile (extracted region marked by the square in **a**).

The wOFV $\varepsilon_u$ sensitivity to $\lambda$ is further evaluated for a broader range of $\lambda$ computed across 118 values shown in Fig. 4. Increments of $\delta\lambda = 1$ are used for the first 20 values to resolve the initial rapid $\varepsilon_u$ variation, starting from $\lambda = 0.001$ to $\lambda = 20$, before changing to a coarser spacing of $\delta\lambda = 10$ for the remaining values. While this $\lambda$ sensitivity is shown for a single image pair, trends are consistent for all image pairs within the 100 image sequence. The selected $\lambda = [2, 40, 100, 180, 520, 1000]$ values discussed throughout this work are shown in Fig. 4. For clarity, these chosen $\lambda$ values correspond to under-regularized ($\lambda = 2$), slightly under-regularized with $\varepsilon_u$ equivalent to PIV ($\lambda = 40$), near minimum $\varepsilon_u$ ($\lambda = 100$), minimum $\varepsilon_u$ ($\lambda^* = 180$), slightly over-regularized with $\varepsilon_u$ equivalent to PIV ($\lambda = 520$), and over-regularized ($\lambda = 1000$). The corresponding PIV error for the same image pair is marked by the red line for comparison.

As shown in Fig. 4, wOFV results for $\lambda < 40$ give unacceptable $\varepsilon_u$ values significantly greater than PIV. The large $\varepsilon_u$ is a result of the noise introduced into the under-regularized flow field. As $\lambda$ increases above 40, $\varepsilon_u$ values decrease less rapidly to the minimum $\varepsilon_u$ at $\lambda^* = 180$. A gradual, linear increase in $\varepsilon_u$ past $\lambda^*$ ensues as the estimated velocity field becomes increasingly over-regularized. For the flow field in Fig. 2, $\lambda$ values in the range $\lambda = 40 - 520$ provide modestly more accurate velocity fields than PIV, at best reaching a ~23% improvement in $\varepsilon_u$ at $\lambda^*$. The exact range of $\lambda$ yielding improvements over PIV varies slightly from image to image. However, it is positive to see a broad range of $\lambda$ yield acceptable error values beyond the current state-of-the-art and shows the strength of the current wOFV approach.

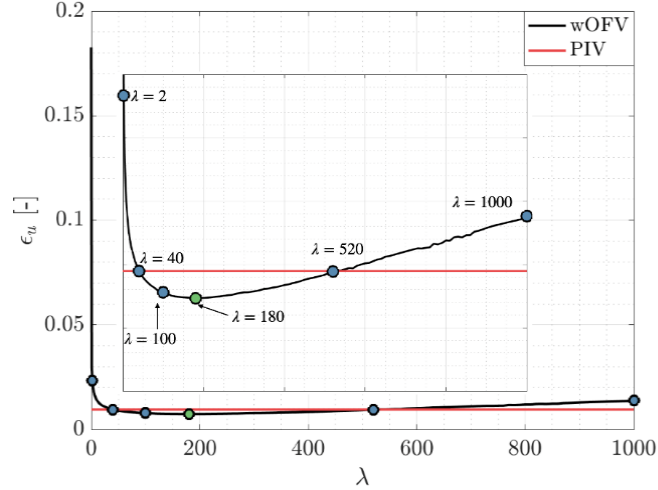

**Fig. 4** Sensitivity of wOFV $\varepsilon_u$ as a function of $\lambda$ for the velocity field in Fig. 2.

### 4.1.2 Image sequence analysis

The findings in Sect. 4.1.1 consider a single image pair from the synthetic dataset. The influence of $\lambda$ for the complete 100 image sequence, which involves a temporally varying wall-bounded flow, will be now considered. Figure 5a shows the $\varepsilon_u$ values for wOFV at selected $\lambda$ values, as well as for PIV across the image sequence. The 100 image average $\langle \varepsilon_u \rangle$ values are shown by the bar chart in Fig. 5b.

For all results presented, the $\varepsilon_u$ for a given image pair can be seen to vary slightly across the image sequence. This variation is dependent on the complexity of the instantaneous flow dynamics for a given image pair as coherent structures and streaks propagate across the image. Despite varying $\varepsilon_u$ values across the image sequence, the trends remain consistent with those presented for the single image pair. In particular, $\varepsilon_u$ values are exceptionally large for the under-regularized $\lambda = 2$ value, but $\varepsilon_u$ decreases substantially as $\lambda$ increases. $\varepsilon_u$ values are lowest for $\lambda^* = 180$, but wOFV findings with $\lambda = 100$ yield similarly low $\varepsilon_u$ values, which is consistent with the broad local minimum curve feature shown in Fig. 4. wOFV findings with $\lambda = 40, 520$ yield comparable $\varepsilon_u$ values as PIV. As $\lambda$ increases beyond 180 values $\varepsilon_u$ gradually increase but remain lower than $\lambda = 2$.

Overall, the error analysis reveals that wOFV can surpass PIV accuracy for a relatively broad range of $\lambda$ values consistent with Fig. 4. wOFV can provide improvements in accuracy up to 24% compared to PIV. In addition, the gradual increase in $\varepsilon_u$ for over-regularized $\lambda$ values compared to the sharp rise in $\varepsilon_u$ for under-regularized $\lambda$ values suggest that in the absence of a ground truth reference, it may be preferable to select over-regularized as opposed to under-regularized velocity estimates. However, it should be emphasized that $\varepsilon_u$ represents a *spatially averaged* value across the *entire* image domain. It is unlikely that a single $\lambda$ value is optimal for all locations of the velocity field. This has already been seen in Fig. 3b, where it was shown that $\lambda$ values closer towards the under-regularized side of $\lambda^*$ were able to resolve sharp velocity gradients compared to $\lambda^*$. This finding is discussed further in the following section.

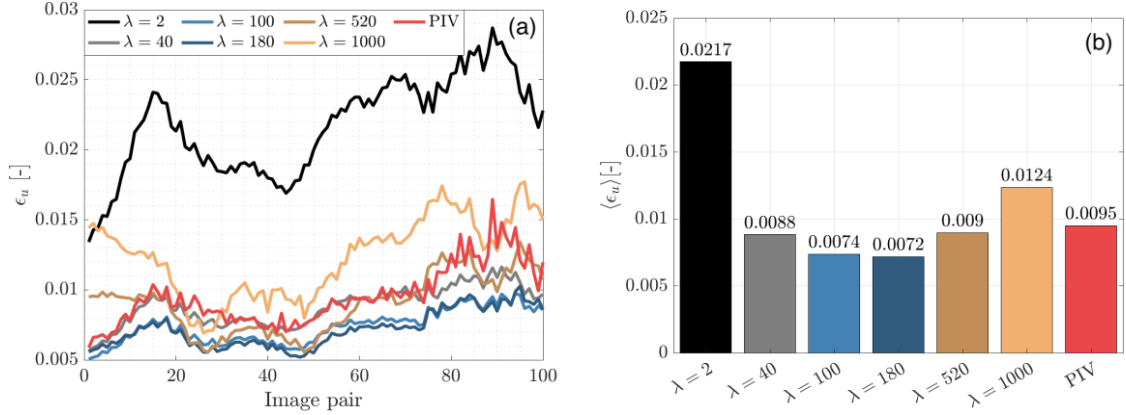

**Fig. 5 a** wOFV and PIV $\varepsilon_u$ across 100 image sequence. **b** 100 image average value.

### 4.1.3 Regional $\lambda$ sensitivity

A single $\lambda$ value weighting applied to an entire image can lead to a non-optimal velocity estimation in various regions of an image. It is thus important to understand the local distribution of error across the individual boundary layer regions. In this section, the error from the ground truth is evaluated within each boundary layer region contained within the synthetic images. This analysis reveals the trend of $\lambda$ to optimize wOFV accuracy in each boundary layer region. For clarity, Fig. 6a shows the physical domain of the viscous sublayer ($y^+ < 5$), buffer layer ($5 < y^+ < 30$) and logarithmic region which covers the remainder of the image field of view ($30 < y^+ < 138$) in this dataset. In addition, the full viscous sublayer resolvable by wOFV is considered in this analysis, as opposed to only considering the equivalent PIV region as performed for $\varepsilon_u$.

The *unnormalized* root mean square error (RMSE) is calculated to quantify the absolute error within each boundary layer region:

$$\text{RMSE} = \sqrt{\frac{1}{n_v}\sum (U_1 - U_{1,DNS})^2 + (U_2 - U_{2,DNS})^2} \qquad (14)$$

In contrast to $\varepsilon_u$, the absence of normalization by the DNS magnitude in the RMSE avoids an exaggeration of errors closest to the wall where the velocity magnitude approaches zero. The 100 image average RMSE in each region is shown in Fig. 6 b-d for the various $\lambda$ values and for PIV.

In Fig. 6, it can be seen that the RMSE trend as a function of $\lambda$ is regionally dependent. In the logarithmic region, wOFV performs exceptionally well for the previously acclaimed over-regularized $\lambda = 520, 1000$, but suffers from high RMSE as the velocity field becomes more under-regularized. In the buffer layer, wOFV is sensitive to both under- and over-regularization; while the under-regularized $\lambda = 2$ still yields the highest RMSE, the RMSE for the over-regularized $\lambda = 520, 1000$ more than doubles compared to the logarithmic region and the optimal $\lambda$ decreases from $\lambda = 520$ to $\lambda = 100$. In the viscous sublayer, wOFV now becomes more sensitive to over-regularization, as $\lambda = 1000$ now has the highest RMSE, while the RMSE for $\lambda = 2$ decreases substantially and the optimal $\lambda$ decreases to 40. PIV performs consistently well in each boundary layer region, however, wOFV at its optimized $\lambda$ values achieves RMSE improvements of 21%, 11% and 29% in accuracy over PIV in the viscous sublayer, buffer layer and logarithmic region, respectively.

This trend of wOFV preferring lower $\lambda$ values and becoming more sensitive to higher $\lambda$ values as the wall is approached can be explained by considering the particular flow dynamics in these regions. In the logarithmic region, velocity gradients are weaker compared to closer to the wall. Therefore, effects of over-smoothing in the logarithmic region will have less of a detrimental effect on accuracy as the motion is predominantly uniform. As the wall is approached in the buffer layer, stronger velocity gradients exist requiring a slightly lower $\lambda$ to resolve them without over-smoothing. In the viscous sublayer, where the lowest velocities are present, even lower $\lambda$ values are preferred to resolve the sub-pixel particle displacements and consistently large velocity gradients, which are both significantly more sensitive to over-smoothing than noise compared to the regions away from the wall.

These findings demonstrate that a single $\lambda$ value can slightly compromise the wOFV accuracy within the various regions of the boundary layer. While spatially adaptive regularization schemes have been proposed in the literature (Stark, 2013; Lu et al., 2021; Ouyang et al., 2021; Zhang et al., 2020), implementation of these schemes is non-trivial and is beyond the scope of this work. Although wOFV cannot be fully optimized using a single $\lambda$ value, these findings positively demonstrate that values between $\lambda = 100 - 180$, including the optimal on average $\lambda^*$, provide well-balanced solutions in each region with wOFV offering up to 23% improved accuracy over PIV.

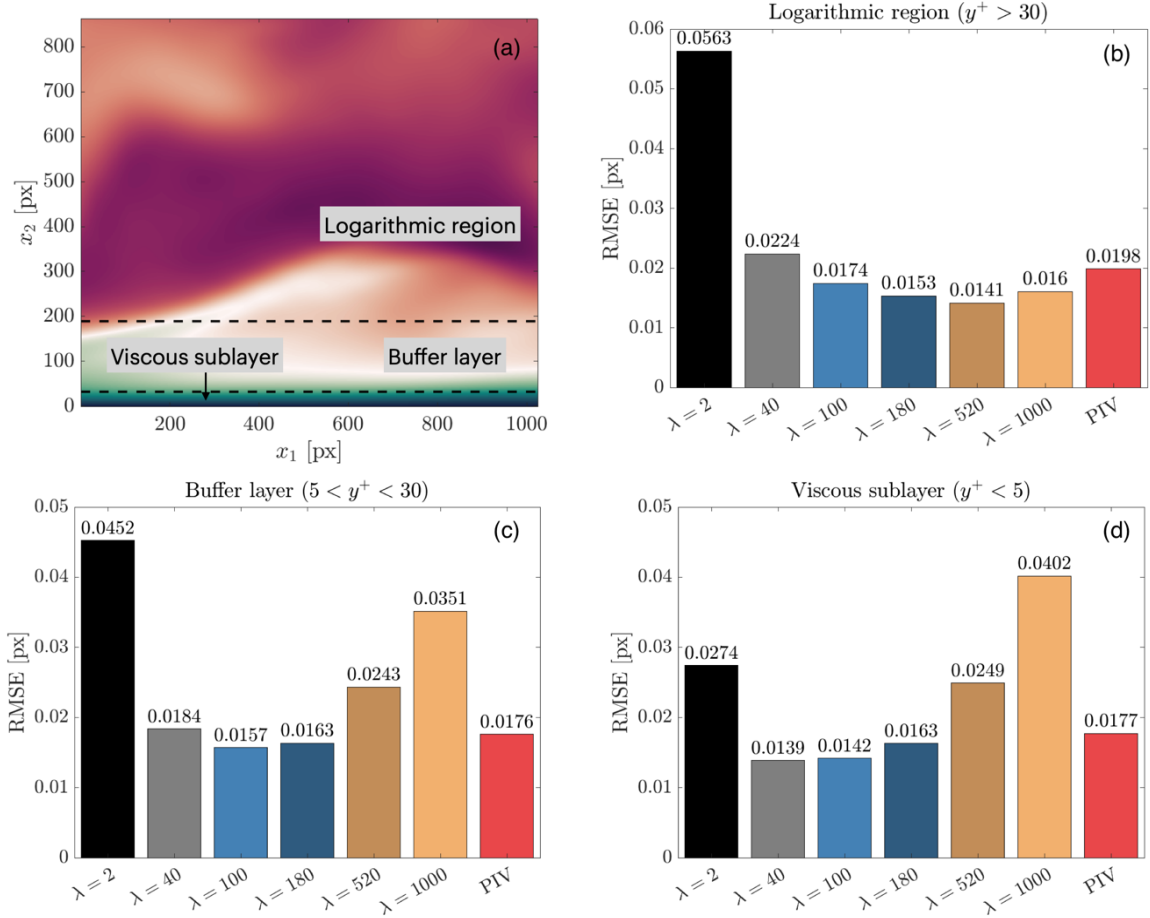

**Fig 6 a** DNS velocity field with marked regions. Variation of average RMSE across image sequence as a function of $\lambda$ for the **b** Logarithmic region, **c** Buffer layer and **d** Viscous sublayer.

### 4.2 $\lambda$ sensitivity in the near-wall region

This section evaluates wOFV's ability to estimate the mean velocity behavior within the boundary layer by analyzing the normalized velocity profiles depicted by $u^+, y^+$. The effect of lambda on wOFV to accurately calculate $\langle U_1 \rangle$ and the near-wall velocity gradient $\gamma$ is first assessed, since these variables are necessary to calculate $u^+$ and $y^+$. Subsequently, the fidelity of wOFV to resolve the various turbulent boundary layer regions is evaluated. The ensemble-mean velocity field presented in this section is composed from 100 velocity images, and is evaluated separately for each $\lambda$ as well as for PIV.

#### 4.2.1 Viscous sublayer mean velocity
The ensemble average streamwise velocity $\langle U_1 \rangle$ within the viscous sublayer ($y^+ < 5$) is shown in Fig. 7a and a zoomed view closest to the wall is shown in Fig. 7b. The $\langle U_1 \rangle$ profiles shown are extracted from the $x_1 = 400$ location marked by the dashed line in Fig. 2. Immediately obvious in Fig. 7 is the finer vector resolution for wOFV compared to PIV; wOFV provides vectors per pixel all to way to the wall, while PIV has one-fourth the vector spacing and resolves approximately half of the viscous sublayer region.

The effect of increasing $\lambda$ on wOFV is evident in Fig. 7b. As $\lambda$ increases, the streamwise velocity approaching the wall is elevated and increasingly deviates from the no-slip condition at the wall as the vector field becomes over-smoothed. For $\lambda \leq 180$, the deviation from the DNS is significantly less with low velocities of $\langle U_1 \rangle = 0.003$ to $0.01$ at the wall. wOFV with $\lambda = 2, 40$ provides the best agreement with DNS, which is consistent with the error distribution analysis in Sect. 4.1.3 showing slightly under-regularized wOFV results produce the most accurate vector estimates in the viscous sublayer. However, the differences between $\lambda = 2, 40$ and $\lambda^* = 180$ are small (< 3%). This smoothing effect of the regularization term $J_R$ becomes particularly more apparent for $\lambda > 180$. This tendency for the regularization term to dominate and oversmooth at motion and image intensity discontinuities is well-known in optical flow literature and is also influenced by using a quadratic penalty in $J_R$ (Zach et al., 2007). Compared to the DNS and wOFV results, PIV estimates a slightly lower velocity within the resolved PIV region down to $x_2 = 14$. Although relatively minor, this systematic error occurring in the vicinity of the wall is absent in all of the wOFV results.

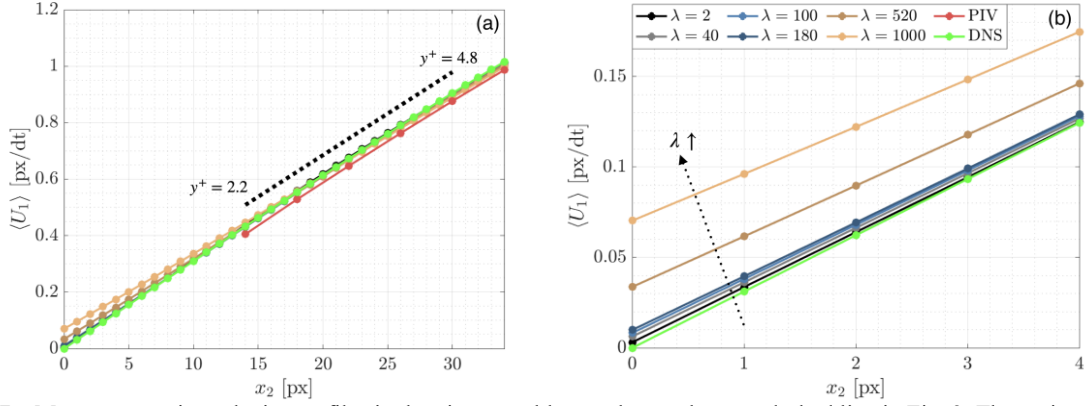

**Fig 7 a** Mean streamwise velocity profiles in the viscous sublayer taken at the grey dashed line in Fig. 2. The region used for the $\gamma$ calculation for PIV is marked by the dashed line. **b** Highlight of the final wOFV vectors

### 4.2.2 Near-wall gradient

Having established how $\lambda$ affects estimates of $\langle U_1 \rangle$ in the vicinity of the wall, it is necessary to understand how these effects propagate into deriving the near-wall gradient $\gamma$ and therefore the friction velocity $u_\tau$ needed for the normalization of boundary layer quantities. Accurate and direct estimation of $\gamma$ can be challenging for several reasons. In particular, there is the need to resolve reliable velocity vectors as close to the wall as possible and maximise the spatial resolution. The sharp velocity gradient also needs to be resolved reliably in the presence of the image discontinuity (i.e., the masked wall region).

The calculation of $\gamma$ is performed by using a linear regression routine. For PIV, linear regression is performed from $y^+ = 4.8$ to the final vector at $y^+ = 2.2$ as illustrated by the dashed line in Fig. 7. For wOFV, linear regression is applied at $y^+ = 4.8$ and extends to $y^+ = 0.32$ to avoid the no-slip pixel at $x_2 = 0$. The regression calculation includes 5 vectors for PIV and 28 vectors for wOFV. A normalized percentage error in $\gamma$ is calculated by:

$$\varepsilon_\gamma = \frac{|\gamma - \gamma_{DNS}|}{\gamma_{DNS}} \times 100 \qquad (15)$$

The true $\gamma_{DNS}$ is calculated using a linear regression across the same respective regions for each technique. The near-wall gradient error $\varepsilon_\gamma$ is calculated at each valid pixel position for wOFV away from the image edges and a subsampled $\gamma_{DNS}$ is used for comparison with the lower resolution PIV grid. The mean average of the normalized near-wall gradient error $\varepsilon_\gamma$ across the available $x_1$ distance is shown in Fig. 8.

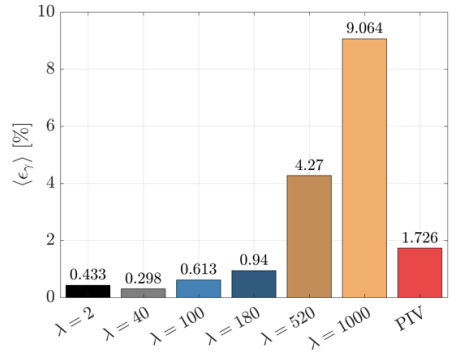

**Fig 8** Average $\varepsilon_\gamma$ error across the $x_1$ distance.

Figure 8 shows that the under-regularized $\lambda$ values are more conducive to reduced error in $\gamma$, with $\lambda = 40$ achieving the minimum $\langle \varepsilon_\gamma \rangle$. This trend is similar to the RMSE within the viscous sublayer (Fig. 6d); however, now the under-regularized $\lambda = 2$ outperforms $\lambda = 100, 180$. wOFV with $\lambda = 2$ performs better for $\varepsilon_\gamma$ than for the RMSE because the noise present in each image at low $\lambda$ is mostly washed out when calculating the ensemble mean velocity $\langle U_1 \rangle$. Despite the preference towards under-regularization, it must be emphasized that wOFV results for $\lambda = 2$ to $\lambda^* = 180$ all provide higher accuracy than PIV. The $\varepsilon_\gamma$ values for this $\lambda$ range remain less than 1% and are a 45%-83% improvement over PIV. In contrast, the over-regularized $\lambda = 520$ and $1000$ cases have serious and unacceptable levels of error, which are 147%-425% greater than PIV. These unacceptable errors are a result of over-smoothing the velocity field at the wall as shown in Fig. 7. Clearly over-regularization should be avoided when evaluating velocity quantities closest to the wall.

### 4.2.3 Normalized Mean Velocity Profile

The normalized $u^+$ velocity profiles are analysed to understand the effect of $\lambda$ on wOFV's ability to interpret the mean streamwise velocity in each region of the boundary layer. The inner scaled profiles are presented in Fig. 9, taken at the location marked by the dashed line in Fig. 2. The relations for the linear $u^+ = y^+$ viscous sublayer and logarithmic region $u^+ = 1/\kappa \ln(y^+) + \beta$ with the constants $\kappa = 0.41$ and $\beta = 5.2$ (Pope, 2002) are indicated by the dashed lines. For the results in Fig. 9, each profile is normalised using its respective $u_\tau$ calculated from $\gamma$.

When considering the results in Fig. 9, it is first important to discuss the effect of $\gamma$ on the velocity profiles. Recall that over-regularized $\lambda = 520, 1000$ yields an underestimation of $\gamma$ due to over-smoothing the velocity field. According to Eqns. 1-4, an underestimated $\gamma$ will decrease $y^+$ and increase $u^+$, creating a slight vertical and leftward shift in the normalized velocity profiles for $\lambda = 520, 1000$. In the viscous sublayer ($y^+ < 5$), this shift, as well as general over-smoothing of $\langle U_1 \rangle$, creates a deviation from DNS and the established linear $u^+ = y^+$ relationship. This shift also causes a mild deviation from the DNS throughout the buffer layer ($y^+ \approx 5 - 30$) followed by a more noticeable deviation in the logarithmic region ($y^+ \approx 30 - 200$). wOFV findings from $\lambda = 2$ to $\lambda^* = 180$, which are not over-regularized, show excellent agreement with DNS throughout each region of the boundary layer. As shown in Fig. 9, the noise associated with the under-regularized $\lambda = 2$ result is mostly washed out when considering the ensemble average velocity $\langle U_1 \rangle$. Although minor fluctuations due to under-regularization can be seen in the logarithmic region for $\lambda = 2$ (see Fig. 9c), these fluctuations are smaller than the deviations present for over-regularized wOFV results.

Figure 9 shows that PIV is broadly in good agreement with the DNS. A slight discrepancy in the logarithmic region exists for PIV, but not to the extent of $\lambda = 520, 1000$. PIV resolves down to a minimum $y^+ = 2.21$ in the viscous sublayer. Excluding the over-regularized $\lambda = 520, 1000$, wOFV resolves 2 decades in wall units further than PIV, down to $y^+ = 0.15$ while maintaining agreement with the DNS with an error less than $0.05 \, \delta u^+$ for the final vector at the wall. Assuming a suitable $\lambda$ is selected, these results show highly encouraging performance characteristics of wOFV in terms of improved accuracy and increased vector density, which enables better interpretation of the viscous sublayer.

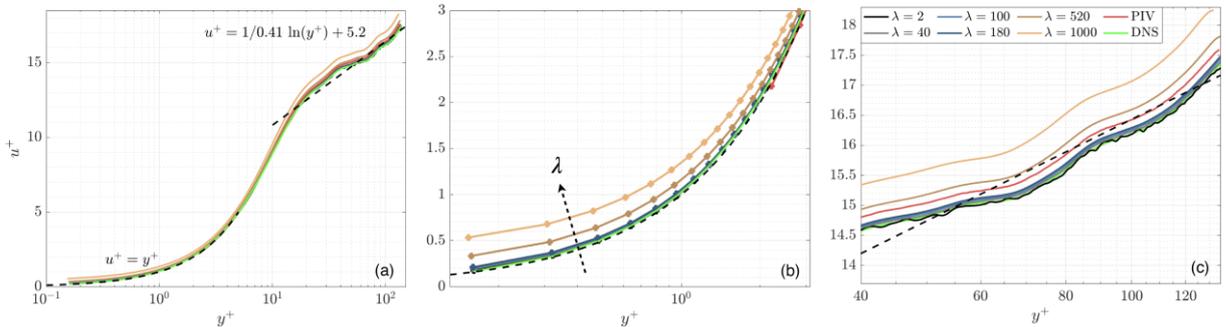

**Fig 9 a** Inner scaled mean velocity profiles. Zoomed regions of the **b** Viscous sublayer and **c** Logarithmic region. Theoretical relations for the viscous sublayer and logarithmic region are shown in the dashed lines.

## 5. Application to experimental data

While synthetic data is key for quantifying and understanding error characteristics of wOFV, it is essential to further evaluate the performance of the method on a real experimental dataset which departs from the simplicity of synthetic data. In Sect. 4, while it was shown that wOFV provides improvements in accuracy over PIV, PIV performed exceptionally well for the synthetic data, which is absent of noise and other imaging artifacts. It should again also be emphasized that the degree of smoothing in the PIV results was optimized for maximum accuracy. This was only possible since the DNS ground truth velocity was available for comparison. True experimental data, on the other hand, does not have such a reference and often suffers from inherent camera noise, laser pulse variation and non-uniform illumination from reflections near the wall, which can present additional difficulties to obtain accurate velocity measurements. While image pre-processing methods can alleviate some of these effects, in practice it is not possible to avoid them entirely.

Applying the knowledge gained from the synthetic data, in this section wOFV is applied to experimental particle images of a developing turbulent boundary layer. wOFV results from a selection of $\lambda$ values are compared to PIV as well as a PIV + PTV approach, which provides higher spatial resolution than PIV. This comparison demonstrates the advantages of wOFV over PIV and PTV to resolve the turbulent boundary layer flow features with improved vector resolution and accuracy.

## 5.1 Experimental Setup

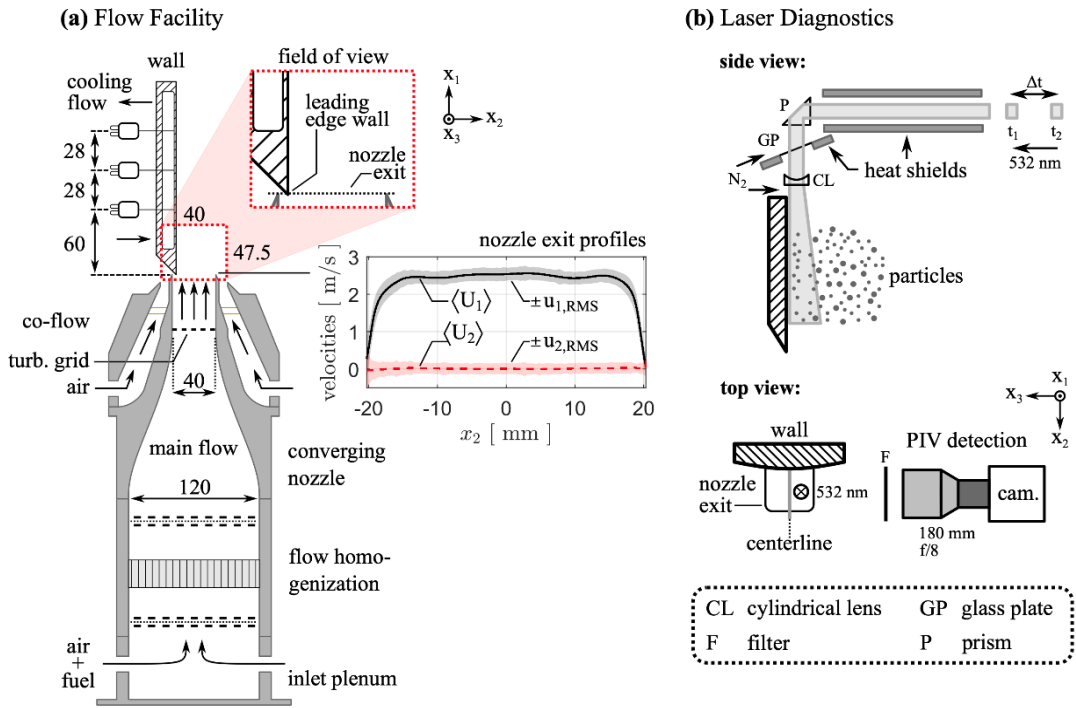

**Fig. 10**: Schematic of **a** Flow facility (SWQ-burner) in a side view **b** Applied laser diagnostics in a side and top view. Numbers without units indicate spatial dimensions in millimeters.

Experiments are conducted in a flow facility in which a jet flow impinges onto a parallel wall, creating a developing turbulent boundary layer. The flow facility was originally designed to study flame-wall interactions in a side-wall quenching (SWQ) configuration (Jainski et al., 2018; Kosaka et al., 2018; Kosaka et al., 2020; Zentgraf et al., 2021; Zentgraf et al., 2022). For the purposes of this study, the flow facility operates under non-reacting, cold-flow conditions (i.e., no combustion). This experimental setup was recently presented in (Zentgraf, 2022) for characterizing the nozzle exit velocity profiles of the SWQ-burner. A schematic of the facility is shown in Fig. 10a. The central main flow (fully premixed $CH_4$/air at $\phi = 1.00$; not ignited) was homogenized by meshes as well as a honeycomb structure and subsequently guided through a converging nozzle. At the square nozzle exit ($\approx 40 \times 40$ mm$^2$) the Reynolds number was maintained at 5900 and the inflow conditioning yielded a streamwise ($x_1$) velocity profile with a nearly top-hat shape (Zentgraf, 2022). For turbulent conditions at the nozzle exit, a turbulence grid was used, providing a turbulence intensity of 6-7% (Jainski et al., 2018). The outlet flow impinged the sharp leading edge of a stainless-steel wall. The wall's surface has a mild curvature for improved optical access (radius: 300 mm, see top view in Fig. 10b). The central main flow was shielded from the lab environment by a concentric square air co-flow. All flows operated at ambient temperature, which was in agreement with the wall temperature.

A low-speed (10 Hz) PIV setup was used as shown in Fig. 10b. This setup was used previously to characterize the velocity profiles at the nozzle exit as boundary conditions (Zentgraf, 2022) and its optical arrangement closely matched the high-resolution, high-speed realization in (Zentgraf et al., 2021). The main flow was seeded with $Al_2O_3$ particles (Zentgraf et al., 2021) which were illuminated using a dual-cavity Nd:YAG PIV laser (New Wave Research, Gemini PIV, G200, 10 Hz, 532 nm). Laser pulses were separated by $\Delta t = 40\ \mu s$. The laser sheets were guided vertically downward to the wall to minimize reflections at the wall. Measurements were taken in the $x_1 x_2$-symmetry plane of the facility, at the wall's centerline ($x_3$ = 0 mm). The origin of the coordinate system is defined at the leading edge of the wall along its centerline. Optics exposed to seeding were continuously purged by nitrogen during operation.

The resulting Mie-scattering was detected by a sCMOS camera (LaVision GmbH, Imager sCMOS) with an exposure time of 15 $\mu$s each frame. The camera was equipped with a 180-mm objective lens (Sigma, APO Macro DG HSM D, f/8) and a bandpass filter (Edmund Optics Inc., #65-216, central wavelength 532 nm, FWHM 10 nm) to suppress ambient light. The field of view (FOV) comprises $(\Delta x_1, \Delta x_2) \approx (40\ \text{mm}, 47.5\ \text{mm})$. For velocimetry, the images are cropped to $(\Delta x_1, \Delta x_2) \approx (38\ \text{mm}, 38\ \text{mm})$, comprised of 2048 × 2048 pixels with the FOV beginning at the wall's leading edge. At the downstream edge of the FOV, the Reynolds numbers based on the momentum thickness and friction velocity are $Re_\theta = 100$, $Re_\tau = 70$.

## 5.2 Vector field calculation

For the experimental dataset, wOFV is benchmarked against PIV as well as PIV + PTV, the latter of which is often used in experiments to improve vector resolution over PIV. It is emphasized that experiments were originally optimized for PIV/PTV. The seeding density was optimized to provide 6-8 particles per final interrogation window and particle displacement was within ¼ of the final interrogation window size in the near-wall region of investigation. Velocity vector fields achieved average cross-correlation values of 0.77. It is therefore emphasized that the PIV quality is not intentionally compromised to exaggerate the advantages of wOFV.

The wOFV and PIV velocity fields were processed similarly to that of the synthetic data in Sect. 4. Mie scattering images were first pre-processed with subtraction of the ensemble minimum image followed by a min-max intensity normalization (Adrian & Westerweel, 2011). PIV vector processing was performed using a multi-pass correlation with an initial IW size of 64 × 64 down to 16 × 16 with 75% overlap. The same anisotropic denoising filter used to optimize the PIV results on the synthetic data was applied to the experimental PIV vector fields. PIV + PTV processing was initialized from PIV. PTV was calculated for a particle size range from 1 – 8 pixels and with a correlation window size of 8 pixels. PTV vectors were converted to a structured 4 × 4 pixels[2] grid, as performed in previous boundary layer studies (Ding et al., 2019; M. Schmidt et al., 2021). This step was performed in DaVis using a "simple averaging / strong filter" scheme in DaVis, which provided the most reliable PTV results. A 3 × 3 Gaussian smoothing filter was applied to remove noise in the PTV vector fields. wOFV was performed as described in Sect. 2.

Each velocimetry method provides a different vector spacing and spatial resolution. For PIV, the spatial resolution is 298 $\mu m$, as defined by the final IW size of 16 × 16, while 75% overlap provides a vector spacing of 74.3 $\mu m$ (every 4 pixels). The converted 4 × 4 pixels[2] grid used for PIV + PTV provides a vector spacing of 4 pixels or 74.3 $\mu m$, which is equivalent to PIV. Since PTV assigns a vector to the centroid of each detected particle, an approximate PIV + PTV spatial resolution is reported as the average particle distance of 5.8 pixels or 107.8 $\mu m$. wOFV provides a per-pixel vector spacing of 18.6 $\mu m$. A conservative estimate of wOFV's spatial resolution is reported as the average particle spacing of 107.8 $\mu m$. As mentioned in Sect. 4, wOFV's true spatial resolution is likely to be smaller than the average particle spacing since each particle pixel contains a valid vector, which likely makes the particle centroid spacing an upper limit.

The first vectors from the wall are located 279 $\mu$m, 204 $\mu$m and 149 $\mu$m for PIV, PIV + PTV and wOFV respectively. These distances are based on geometric masks used to calculate vector fields that are offset from the wall location to avoid light reflections and reduce the frequency of spurious vectors at the wall for both methods. The wall location is approximated using the maximum intensity of the reflection present at the wall. This is then refined using the no-slip pixel position estimate from PIV and wOFV $\lambda = 0.1 \langle U_1 \rangle$ profiles averaged over the downstream distance.

It should be emphasized that the experimental dataset is appreciably different from the synthetic dataset in Sect. 4, and this influences the optimization of wOFV. For example, the near-wall velocity gradient $\gamma$ is larger and the viscous sublayer is much thinner for the experimental data than the synthetic data. In addition, the image size is $2048 \times 2048\ px$ compared to $1024 \times 1024\ px$ in the synthetic data. Therefore, not only is the absolute pixel-wise length halved, but the viscous sublayer comprises a smaller and less significant proportion of the full image FOV. Lastly, while the synthetic data has an average freestream particle displacement of 3 pixels, the experimental freestream flow field has a more substantial displacement of 6 pixels. All of these aspects, in addition to different image characteristics, will influence the regularization weighting for wOFV, such that a suitable range of $\lambda$ values will be significantly different between the experimental and synthetic datasets. This aspect is common within optical flow literature (Kadri-Harouna et al., 2013). In fact, the experimental data are significantly stricter and less forgiving compared to the synthetic data regarding selection of an acceptable $\lambda$. The absence of ground truth data means determining the true optimal $\lambda$ is not possible. For the experimental data, wOFV results from three values of $\lambda = 0.1, 1, 20$ are presented and the most appropriate $\lambda$ is justified *a posteriori* based on physical principles as well as general comparison to PIV.

### 5.3 Instantaneous velocity fields

Assessment of wOFV first considers the instantaneous velocity field in comparison to PIV and PIV + PTV. Figure 11 shows an instantaneous velocity magnitude field for PIV, PIV + PTV, and the three wOFV results. An insert is shown for each image, which highlights details of a low-speed streak emerging from the wall.

PIV performs a good job resolving the overall velocity field. However, PIV can often struggle to resolve the velocity near the wall as shown by the pockets of unresolved velocity regions near the wall. PIV + PTV resolves closer to the wall than PIV but resembles a noisier velocity field with similarity to the noise seen in $\lambda = 0.1$. It is noted that converting PTV to a larger grid size of $8 \times 8$ pixels$^2$ did not reduce the noise level in the PIV + PTV.

All three wOFV results resolve similar general features as PIV, but the quality of the velocity field is determined by the choice of $\lambda$. wOFV with $\lambda = 0.1$ exhibits the high-frequency, speckle-like noise commonly associated with highly under-regularized findings. While $\lambda = 0.1$ resolves much of the same larger scale features as PIV and PIV + PTV, several artefacts of locally higher and lower velocities exist throughout the image. For wOFV with $\lambda = 1$, the high-frequency noise is removed and the velocity field has strong agreement with PIV and PIV + PTV. The primary differences between $\lambda = 1$ and PIV is that wOFV does not have spurious or missing vectors near the wall and wOFV resolves velocities closer to the wall. In addition, $\lambda = 1$ does not contain the speckle-like noise presented in PIV + PTV. For $\lambda = 20$, the larger flow features are well captured, but the finer scale features present in PIV, PIV + PTV, and $\lambda = 1$ are mostly removed, likely due to over-smoothing.

The fact that noticeable changes in the velocity field occur over a significantly smaller $\lambda$ range confirm the challenge in selecting the appropriate $\lambda$ for the experimental dataset. While it is not possible to determine a $\lambda$ value that provides the highest accuracy, it would appear that $\lambda = 0.1$ is too under-regularized and $\lambda = 20$ is likely over-regularized. Further analysis of the findings within the turbulent boundary layer is performed to evaluate these aspects and to determine the suitability of $\lambda = 1$.

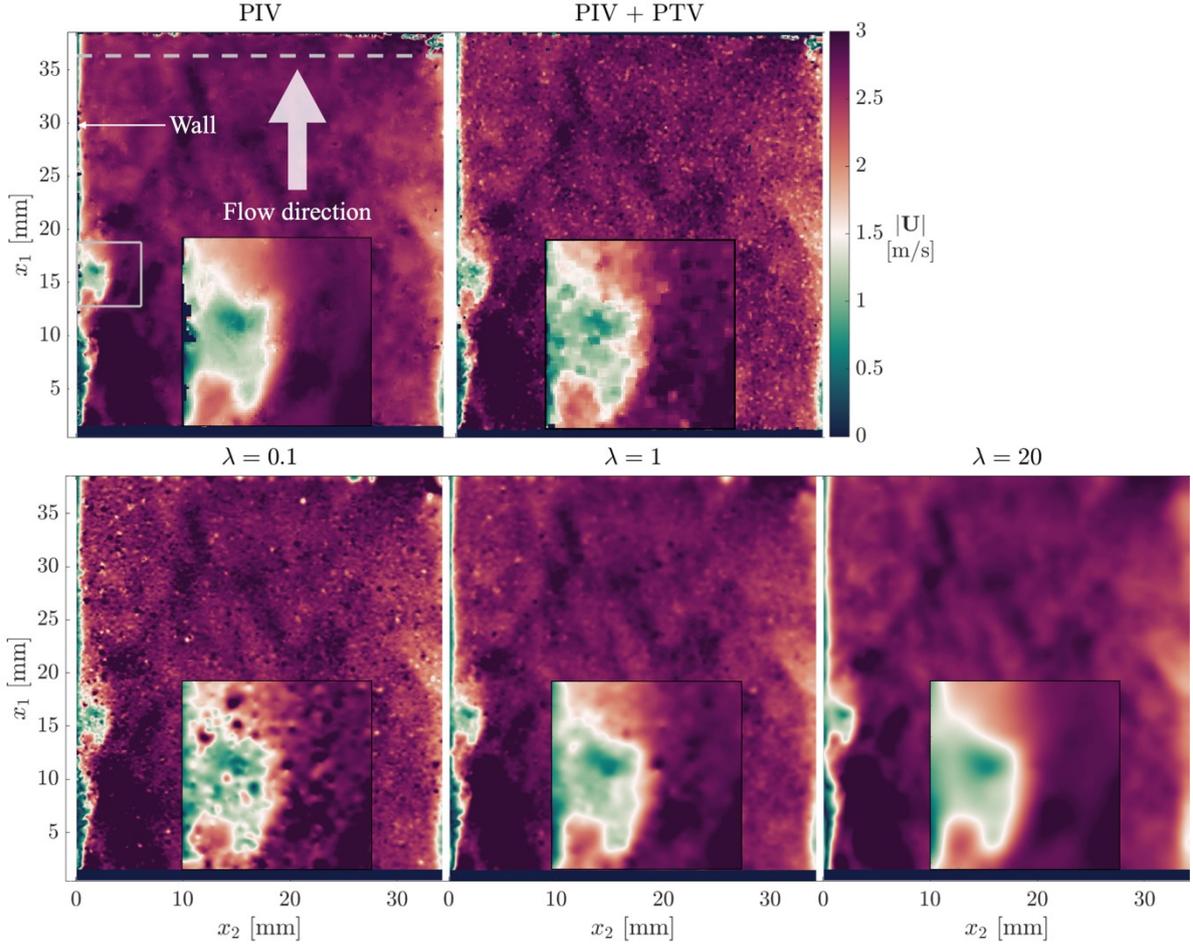

**Fig. 11** Instantaneous velocity magnitude fields. The insert shows a low-speed velocity streak emanating from the wall. The gray dashed line denotes the location where velocity profiles are extracted and analyzed in Fig. 12.

### 5.4 Mean velocity profiles

The near-wall velocity profiles are shown in Fig. 12a, with the normalized profiles shown in Fig. 12b. The $\langle U_1 \rangle$ values are produced from a 100 image mean and the profiles are spatially averaged over a 2mm streamwise $x_1$ distance centered at the location marked by the gray dashed line in Fig. 11. The near-wall gradient $\gamma$ is calculated from the ensemble average streamwise velocity fields using a linear regression in a similar manner to that described in the Sect. 4.2.2. In the viscous sublayer ($y^+ < 5$), there are 15 velocity vectors for wOFV compared to 3 velocity vectors for PIV and 4 for PIV + PTV. The linear regression for wOFV and PIV + PTV uses each of the available vectors, while for PIV only 2 out of the 3 available vectors are used since the final PIV vector nearest to the wall is frequently spurious. As described in Sect. 4.2, the estimation of $\gamma$ has a direct effect on normalized wall units through $u_\tau$. The $\gamma$ values calculated are $\gamma_{PIV} = 1919$, $\gamma_{PTV} = 2402$, $\gamma_{\lambda=0.1} = 2469$, $\gamma_{\lambda=1} = 2288$, $\gamma_{\lambda=20} = 1701$ 1/s, which provide the corresponding $u_\tau$ values $u_{\tau,PIV} = 0.1697$, $u_{\tau,PTV} = 0.1898$ $u_{\tau,\lambda=0.1} = 0.1924$, $u_{\tau,\lambda=1} = 0.1853$, $u_{\tau,\lambda=20} = 0.1597$. Incorrect estimates of $\gamma$ can result in a strong offset from the exact $u^+ = y^+$ formulation shown by the green dotted line in Fig. 12b. Comparison with this linear relation will be used as an approximate measure to judge the quality of the near-wall vectors in the absence of a ground truth velocity.

In Fig. 12a, the profiles above 1mm from the wall are in excellent agreement. For $x_2 < 1mm$, the $\lambda = 20$ profile shows increasing deviation from all other profiles as the wall is approached. In particular, velocity gradients are weaker leading to a flatter curve and significantly higher velocities at the wall. These features clearly indicate that $\lambda = 20$ is over-regularized; the excessive smoothing washes out the velocity gradient at the wall, creating an underestimate of $u_\tau$. The resulting normalization creates a strong deviation from $u^+ = y^+$ as shown in Fig. 12b, and demonstrates that $\lambda = 20$ is not appropriate since the $\gamma$ estimation is compromised.

In Fig. 12a, good agreement is shown between $\lambda = 0.1, 1$, PIV, and PIV + PTV until $x_2 < 0.4\ mm$, where PIV shows a milder gradient for $0.3 \leq x_2 \leq 0.4\ mm$ followed by a sharper velocity gradient at the last

PIV data point. As will be shown, the last PIV data point is often erroneous, which biases the interpreted flow behavior. In Fig. 12b, PIV is offset from the $u^+ = y^+$, with an abnormal deviation in the curve for the last data point. PTV stays in closer agreement with $\lambda = 0.1, 1$ and is able to resolve closer to the wall than PIV although not to the same extent as wOFV. The resulting normalization to inner variables results in significantly closer alignment with $u^+ = y^+$ for PIV + PTV, although a slight offset remains. The $\lambda = 1$ result, on the other hand, shows perfect alignment with the $u^+ = y^+$ relation and remains in good agreement with a discrepancy of 0.04 $\delta u^+$ at the final vector. This suggests that $\lambda = 1$ provides accurate velocity estimates near the wall, as well as an accurate $\gamma$ estimate. This also indicates that PIV, and to a lesser extent PTV, struggles to correctly estimate $\gamma$ causing a slight shift in the normalized velocity profile, but not to the same extreme as $\lambda = 20$. The $\lambda = 0.1$ result exhibits the highest $\gamma$ at the wall, creating a down- and rightward shift in the normalized velocity profile. This shift was not seen for the under-regularized values in the synthetic dataset, which further emphasizes the higher sensitivity of $\lambda$ for the more challenging experimental dataset compared to the synthetic data.

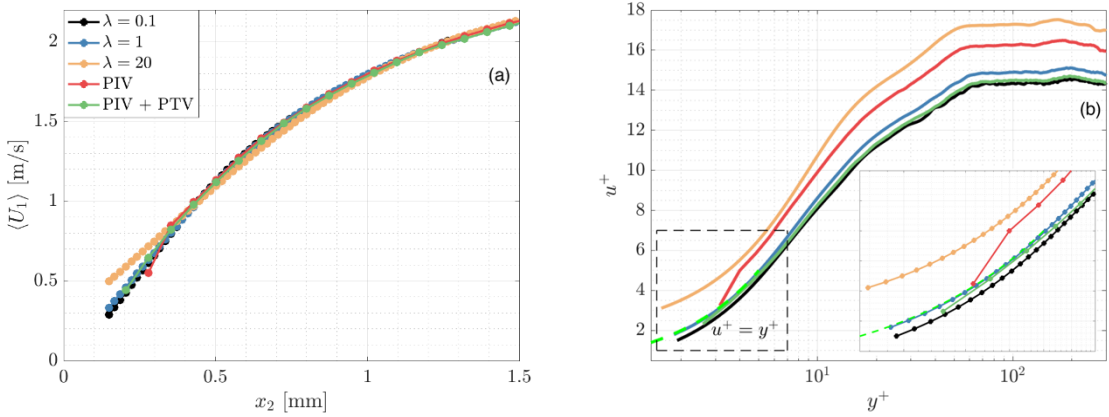

**Fig 12 a** Mean streamwise velocity profiles, **b** Inner scaled mean profiles. Profiles are spatially averaged over across 2mm streamwise $x_1$ distance at the location marked by the grey dashed line in Fig. 11.

## 5.5 Normalized velocity fluctuations

The turbulent velocity fluctuations $u_1 = U_1 - \langle U_1 \rangle$ are analyzed to further evaluate the capabilities of the velocimetry techniques. Velocity fluctuations provide an assessment of the data quality beyond the ensemble mean and are equally important to evaluate turbulent quantities in the boundary layer. Figure 13 shows the profile of the normalized streamwise velocity fluctuations $\langle u_1 u_1 \rangle^+$. The fluctuations and wall-normal coordinate in Fig. 13 are normalized by the $u_\tau$ estimated from $\lambda = 1$ since this $u_\tau$ value provided the strongest agreement with $u^+ = y^+$. Normalizing each case by $u_{\tau,\lambda=1}$ removes the biased curve shifts as shown in Fig. 12. Similar to $\langle U_1 \rangle$, the fluctuation profiles are spatially averaged across the 2mm streamwise distance with the extent of a single standard deviation of these fluctuations illustrated by the shaded area in Fig. 13. The standard deviation of the fluctuations within this 2 mm distance can be considered indicative of the reliability of the velocity estimate and its susceptibility to error.

In Fig. 13, each curve follows a relatively similar trend from $y^+ = 200$ to $y^+ = 10$; $\langle u_1 u_1 \rangle^+$ values increase from the freestream region and exhibit a local maximum in the buffer layer at $y^+ \approx 10$ as seen in other boundary layer studies (e.g., (Spalart, 1988)). From $y^+ = 10$ towards the wall, each curve exhibits different trends. For PIV, $\langle u_1 u_1 \rangle^+$ values continue to increase quite substantially into the viscous sublayer. This trend is non-physical as the turbulent fluctuations are expected to decrease in the viscous sublayer as the wall is approached. PIV also exhibits a very large standard deviation below $y^+ = 10$, which is primarily caused by spurious vectors within the last 2-3 PIV vectors. This feature illustrates the challenges of PIV to accurately resolve small-scale fluctuations in the presence of strong velocity gradients. Reliable PIV measurements are often challenging directly near surfaces. While ensemble-average PIV quantities can be represented with sufficient accuracy, higher order velocity statistics and instantaneous velocity fields more clearly reveal challenges with PIV. PIV + PTV shows improvement from PIV; PTV resolves a greater extent of the buffer layer peak and initially shows the expected decrease in $\langle u_1 u_1 \rangle^+$ towards the wall. However, PIV + PTV still shows the non-physical increase in $\langle u_1 u_1 \rangle^+$ within the final 2-3 vectors at the wall and contains a large standard deviation. Although these artifacts are less severe compared to PIV, they demonstrate that PIV + PTV can still struggle to accurately resolve the flow nearest the wall.

wOFV findings, on the other hand, do not exhibit such large deviations in $\langle u_1 u_1 \rangle^+$, indicating that wOFV is less susceptible to the same errors as PIV and PIV + PTV near the wall. Indeed, $\langle u_1 u_1 \rangle^+$ values are large for $\lambda = 0.1$ due to the results being under-regularized; however, $\langle u_1 u_1 \rangle^+$ values and their deviation are significantly lower than those for PIV or PIV + PTV near the wall. Below $y^+ = 10$, all wOFV findings show the expected decrease in $\langle u_1 u_1 \rangle^+$. The profile $\lambda = 20$ shows a milder peak near $y^+ = 10$ and a milder decrease near the wall compared to the other wOFV findings. The $\lambda = 20$ velocity is over-regularized for which excessive smoothing reduces the variation between peak and trough in the curve from $y^+ = 10$ to $y^+ = 1$. $\langle u_1 u_1 \rangle^+$ values for $\lambda = 1$ show the greatest decrease as the wall is approached, which follows the expected trend of turbulent fluctuations being suppressed within the viscous sublayer in close proximity to the wall. In addition, the extent of the shaded region for wOFV remains constant near the wall suggesting that velocity errors are not being influenced by the proximity of the wall. Overall, wOFV with $\lambda = 1$ shows the most promising findings in terms of ensemble-average values as well as behavior of the velocity fluctuations.

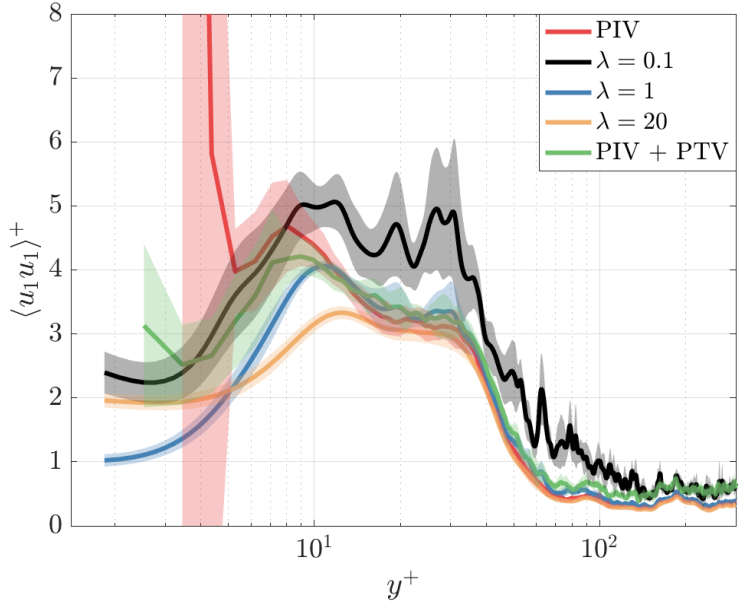

**Fig. 13** Streamwise turbulent fluctuations normalized by $u_{\tau,\lambda=1}$. The shaded regions indicate one standard deviation of the $\langle u_1 u_1 \rangle^+$ values within the 2 mm region centered by the gray dashed line shown in Fig. 11.

### 5.6 Vorticity and turbulent flow structure

An example is presented which highlights the advantages of wOFV in resolving turbulent flow phenomena within a boundary layer. This example is demonstrated for an instantaneous velocity field comparing the optimized wOFV with $\lambda = 1$, PIV and PIV + PTV.

One of the added benefits of wOFV over PIV or PIV + PTV is the improved vector spacing together with physically-sound smoothing, and with that, the ability to better resolve velocity gradient quantities. Figure 14 shows the instantaneous vorticity field $\omega$ for PIV, PIV + PTV, and wOFV. The vorticity is calculated using the 8-point circulation approach described in (Raffel et al., 2018). Individual turbulent structures with relatively high vorticity magnitude are generated near the wall's leading edge and are advected downstream within the developing boundary layer. The inlays shown in Fig. 14 highlight a region that captures a prograde vortex that was generated from the wall's leading edge. This is a particularly challenging region because of the vortex' proximity to the wall, where small pixel displacements coupled with the spatially varying sharp velocity gradients present difficulties to velocimetry techniques. Indeed, PIV has been used to resolve small scale vortex structures near walls, but this is often accomplished by using high image magnifications yielding FOVs smaller than 5x5 mm$^2$ (Jainski et al., 2013), rather than a large FOV that is present in the current work.

The overall vorticity fields calculated from the velocimetry techniques are in good agreement; all methods show similar overall features such as the high vorticity regions extending from the wall's leading edge. PIV + PTV shows higher fluctuations in the vorticity field than PIV and wOFV. This attribute is no doubt due to the higher degree of speckle-like noise present for PTV as shown in Fig. 11. The inserts in Fig. 14 highlight the capabilities of resolving the finer vorticity structures near the wall for each method. Overall, the same spatial distribution of positive/negative vorticity structures are captured by each method, however, the effect of greater vector resolution is immediately seen; in particular, PIV and PIV + PTV images are substantially more pixelated compared to wOFV. PIV can exhibit larger discontinuities in the vorticity field (i.e., larger changes from pixel-to-pixel), which are absent in the PIV + PTV and wOFV results, with wOFV achieving a highly resolved and more continuous vorticity field. The PIV + PTV vorticity field deviates more substantially from PIV and wOFV with several strands of high vorticity extending from the larger vorticity structures. These elevated vorticity strands are likely due to elevated noise levels present in PIV + PTV as discussed in Fig. 11 and 13. wOFV is able to resolve the vorticity much closer to the wall and without troublesome unresolved regions from erroneous vector calculation as in the PIV and PIV + PTV fields. wOFV faithfully preserves the features shown in both the PIV and PIV + PTV results, but achieves a much finer-detailed vorticity field.

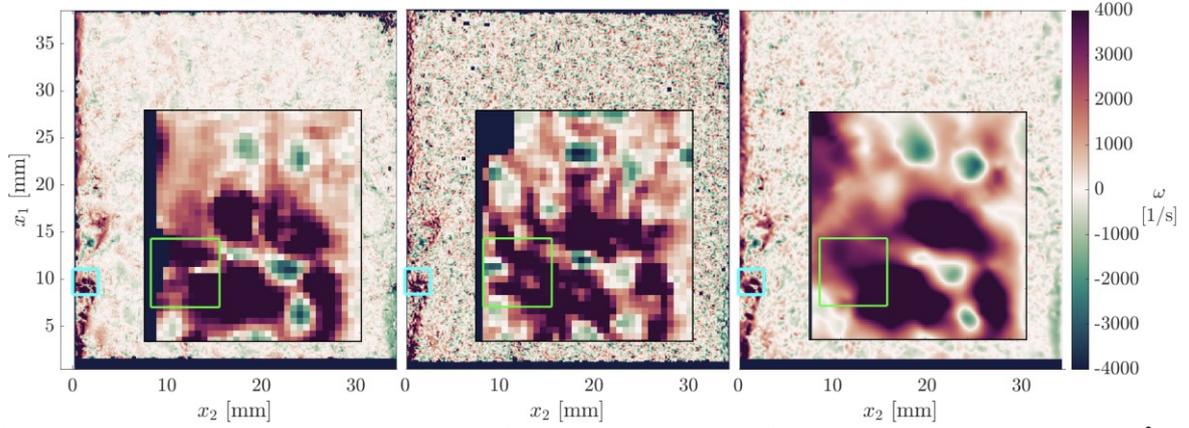

**Fig. 14** Instantaneous vorticity calculated for PIV (left), PIV + PTV (middle) and wOFV (right). Inlays show a 0.8x0.8 mm$^2$ zoomed view of a vortex. The green rectangle indicates the location of the velocity vector field shown in Fig. 15.

Figure 15 shows the corresponding vector field within the green rectangle shown in Fig. 14. The vector field shows all available vectors for PIV, PIV + PTV and wOFV shown in red, blue and black, respectively. wOFV is capable of resolving the prograde vortex in much more detail than the other methods. While the vortex is visible in PIV and PIV + PTV, the vortex structure is more difficult to interpret due to sparser vector spacing and the presence of quasi-erroneous vectors that deviate from a vortical flow pattern. In Fig. 15, all vector fields show good agreement above $x_2 = 0.6\ mm$; most vectors are in alignment and are of the same magnitude. However, closer to the wall there are larger disagreements between wOFV and PIV. In many locations, PIV vectors are aligned orthogonally to wOFV vectors. Some PIV vectors are clearly erroneous as they differ significantly from their neighboring PIV vectors. Additionally, PIV vectors are absent in the upper left corner where spurious vectors are detected and removed during post-processing. Closest to the wall, PIV vectors point inwards towards the wall with a relatively large velocity magnitude, which strongly disagree with the wOFV vectors directed parallel or outward from wall with a velocity magnitude more consistent with the neighboring vectors. PIV + PTV improves on PIV in this regard with suitable quality vectors at the wall and calculates vectors in all regions. However, PIV + PTV exhibits select vectors that disagree with PIV and wOFV. In addition, PIV + PTV vectors near its vortex core center are misaligned with its circulation and struggle to resemble a coherent vortex core. It is likely that PIV + PTV struggles to successfully resolve the strong gradients present in this region. The velocity field features shown in Fig. 15 reveal some challenges cross-correlation-based PIV and combined PIV + PTV experience in resolving small-scale intricate flow dynamics with high velocity gradients in the vicinity of physical boundaries. Assuming a suitable $\lambda$ is selected, these findings positively indicate that wOFV is better suited to resolve these turbulent flow structures in the boundary layer region.

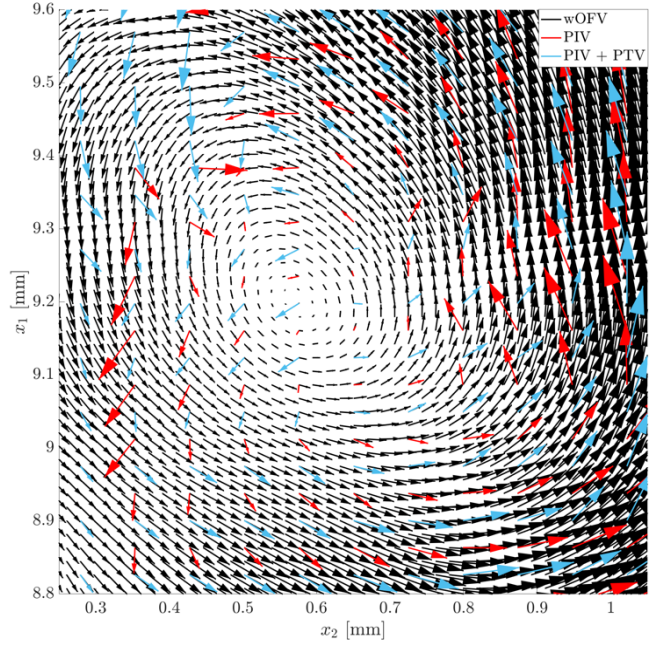

**Fig. 15** Vector field for PIV (red), PIV + PTV (blue) and wOFV (black) within the green rectangle shown in Fig. 14. Vector fields are shown at their original sampling resolutions.

### 5.7 Turbulent energy spectra

Lastly, to assess the potential of resolving fine-scale turbulent velocity fluctuations using wOFV, the normalized streamwise turbulent kinetic energy spectrum ($E_{11}^*(\kappa_1)$) is analyzed. This is calculated using the Fourier transform of the streamwise velocity fluctuations ($u_1$) across the entire field of view. The 1D turbulent kinetic energy spectrum, normalized by its peak value, is presented in Fig. 16 for PIV, PIV + PTV and the optimized wOFV with $\lambda = 1$. Due to the moderately low turbulence level, there is insufficient separation of scales to produce a significant inertial subrange (-5/3 region). The spectra reveal a high frequency noise present for PIV at increasing wavenumbers. The PIV spectra do not show the classical energy decay at increasing wavenumbers, indicating the velocity measurement noise floor and spectral resolution limit has already been reached. The PIV + PTV spectra do not show the high frequency noise present in the PIV profile. However, PIV + PTV spectra show elevated energy at all wavenumbers compared to PIV together with a non-physical modulation after $\kappa_1 > 2 \times 10^4$. wOFV is in close agreement with PIV and PIV + PTV at the low wavenumbers, but shows an energy decay at higher wavenumbers and resolves a significantly greater proportion of the energy spectrum without obvious indications that the measurement is being corrupted by noise or accuracy issues at high wavenumbers.

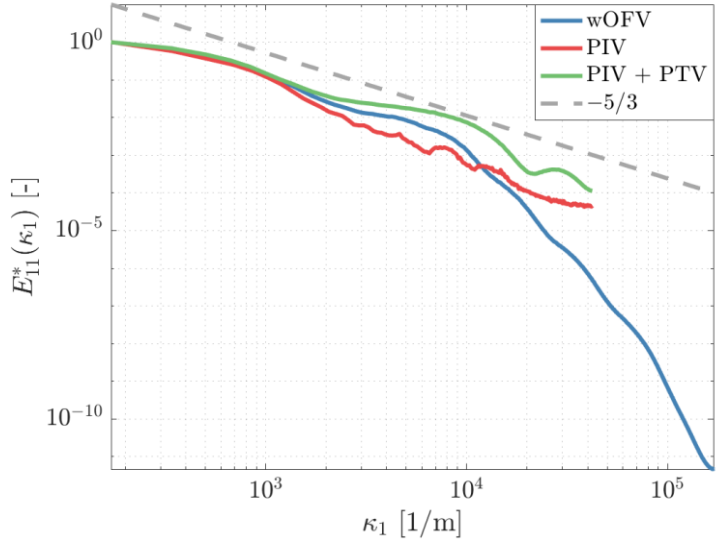

**Fig. 16** Normalized streamwise turbulent kinetic energy spectrum for PIV, PIV + PTV and wOFV.

**Conclusions**

The performance of a wavelet-based optical flow velocimetry (wOFV) method was assessed in detail on synthetic and experimental particle images of turbulent wall-bounded flows. The ability to extract high-resolution estimates of instantaneous, mean and derived flow properties was evaluated in the vicinity of the wall. This was analyzed in regards to selection of the regularization parameter $\lambda$, an aspect largely not discussed in other OFV works, and compared to results from correlation-based PIV.

Using synthetic PIV data generated from DNS of a turbulent boundary layer channel flow, a $\lambda$-sensitivity analysis was performed over the entire field-of-view to establish a range of under-regularized, over-regularized and optimal wOFV results. A regional $\lambda$-sensitivity was investigated to understand the localized error behaviour and considerations necessary to optimize wOFV within each region of the boundary layer. Away from the wall in the logarithmic layer, wOFV is more sensitive to under-regularization, which introduces non-physical noise into the otherwise uniform velocity field. This noise causes significant deviation from the ground truth, leading to unacceptable errors nearly three times greater than PIV. The logarithmic region is less sensitive to over-regularization, since over-smoothing imposed by over-regularization removes noise and produces little deviation to the uniform velocity field. In the buffer layer, wOFV is sensitive to both under- and over-regularization. Over-regularization becomes problematic because over-smoothing washes out velocity gradients present in the buffer layer. In the viscous sublayer, wOFV performs optimally when slightly under-regularized, which better resolves the velocity gradients at the wall in addition to sub-pixel particle displacements. In contrast, over-regularization yields the highest errors as it underestimates the near-wall velocity gradient ($\gamma$). This latter aspect is important when evaluating wall units ($u^+, y^+$) since an underestimated $\gamma$ directly yields an over-estimated $u^+$ and under-estimated $y^+$. Although wOFV vectors at all locations cannot be optimal using a single $\lambda$ value, results confirm a suitable range of $\lambda$ values exist that outperform PIV in each boundary layer region with wOFV also achieving significant improvement in resolving the viscous sublayer more effectively.

The accuracy and resolution improvement is more pronounced when wOFV is applied to experimental images. Physically motivated selection of $\lambda$ based on the expected linear relationship in the viscous sublayer allowed for wOFV to better resolve the mean velocity closer to the wall and stay in excellent agreement with $u^+ = y^+$ down to the final vector. wOFV further provided impressive vector resolution offering 15 vectors in the viscous sublayer, as opposed to PIV and PIV + PTV which respectively offered 3 and 4 vectors in the viscous sublayer with the last vector often being erroneous for PIV. Although PIV performed acceptably when resolving the mean velocity near the wall, evaluation of higher-order velocity statistics and instantaneous flow fields revealed the lower reliability of PIV near walls. In particular, estimates of the turbulent velocity fluctuations from PIV featured a non-physical increase near the wall with unreasonably high standard deviation for the last three vectors closest to the wall. PIV + PTV improved upon such errors, but still exhibited the non-physical increase in turbulent velocity fluctuations and large standard deviation near the wall. wOFV did not exhibit these artifacts. Instantaneous velocity fields further demonstrate the spurious velocity estimations at the wall with PIV. While PIV + PTV exhibited less spurious velocity estimations, noise levels were comparable to the under-regularized wOFV findings, which made it more difficult for PIV + PTV to provide reliable vorticity fields. wOFV does not yield such erroneous velocity estimates, which, together with the improved spatial resolution, allowed for more accurate estimates of derivative quantities detailing complex flow structure in the vicinity of the wall. These findings positively indicate that wOFV is well-suited to estimate the flow dynamics in the presence of physical boundaries.

The authors point out that the wOFV algorithm does not feature direct modifications or explicit constraints for handling physical boundaries within the image. It is expected that such enhancements, although beyond the scope of the current work, would bring further improvement to results and enhance the techniques' performance for velocimetry in more complex geometries.

**Declarations**

**Authors' contributions**

AN implemented the wOFV code, setup the synthetic test case and performed data analysis. The flow experiment was setup and measurements taken by FZ with PIV/PIV + PTV vector processing jointly done by FZ and AN. BP, ML, AD contributed acquisition of funding, technical expertise, supervision and reviewing. All authors were involved in the preparation of the manuscript.

**Ethics approval and consent to participate** Not applicable

**Consent for publication** Not applicable

**Availability of data and materials** Not applicable

**Competing interests**

The authors have no competing interests to declare.

**Acknowledgements**

Funding for wOFV from the European Research Council (grant #759456) and Engineering and Physical Science Research Council (EP/V003283/1) is gratefully acknowledged. Funding for PIV and the experimental setup from the Deutsche Forschungsgemeinschaft (DFG, German Research Foundation) - Projektnummer 237267381 - TRR 150 is also gratefully acknowledged.

**References**

Adrian, R. J., Meinhart, C. D., & Tomkins, C. D. (2000). Vortex organization in the outer region of the turbulent boundary layer. *J. Fluid Mech*, *422*, 1-54. https://doi.org/10.1017/S0022112000001580

Adrian, R. J., & Westerweel, J. (2011). *Particle image velocimetry*. Cambridge university press.

Akima, H. (1974). A Method of Bivariate Interpolation and Smooth Surface Fitting Based on Local Procedures. *Commun. ACM*, *17*(1), 18–20 , numpages = 13. https://doi.org/10.1145/360767.360779

Aubert, G. a. D. R. a. K. P. (1999). Computing Optical Flow via Variational Techniques. *SIAM Journal on Applied Mathematics*, *60*(1), 156-182. https://doi.org/10.1137/S0036139998340170

Barron, J. L., Fleet, D. J., & Beauchemin, S. S. (1994). Performance of optical flow techniques. *International Journal of Computer Vision*, *12*(1), 43-77. https://doi.org/10.1007/BF01420984

Beauchemin, S. S., & Barron, J. L. (1995). The Computation of Optical Flow. *ACM Computing Surveys (CSUR)*, *27*(3), 433-466. https://doi.org/10.1145/212094.212141

Berrut, J. P., & Trefethen, L. N. (2004). Barycentric Lagrange interpolation. *SIAM Review*, *46*(3), 501-517. https://doi.org/10.1137/S0036144502417715

Beylkin, G. (1992). On the representation of operators in bases of compactly supported wavelets. *SIAM Journal on Numerical Analysis*, *29*(6), 1716-1740. https://doi.org/10.1137/0729097

Black, M. J., & Anandan, P. (1996). The robust estimation of multiple motions: Parametric and piecewise-smooth flow fields. *Computer Vision and Image Understanding*, *63*(1), 75-104. https://doi.org/10.1006/cviu.1996.0006

Cai, S., Mémin, E., Dérian, P., & Xu, C. (2018). Motion estimation under location uncertainty for turbulent fluid flows. *Experiments in Fluids*, *59*(8), 1-17. https://doi.org/10.1007/s00348-017-2458-z

Cai, S., Zhou, S., Chao Xu, & Gao, Q. (2019). Dense motion estimation of particle images via a convolutional neural network. *Experiments in Fluids*, *60*, 1-16. https://doi.org/10.1007/s00348-019-2717-2

Chen, L. F., Liao, H. Y. M., & Lin, J. C. (2002). Wavelet-based optical flow estimation. *IEEE Transactions on Circuits and Systems for Video Technology*, *12*(1), 1-12. https://doi.org/10.1109/76.981841

Corpetti, T., Heitz, D., Arroyo, G., Mémin, E., & Santa-Cruz, A. (2006). Fluid experimental flow estimation based on an optical-flow scheme. *Experiments in Fluids*, *40*(1), 80-97. https://doi.org/10.1007/s00348-005-0048-y

Corpetti, T., Mémin, É., & Pérez, P. (2002). Dense estimation of fluid flows. *IEEE Transactions on Pattern Analysis and Machine Intelligence*, *24*(3), 365-380. https://doi.org/10.1109/34.990137

De Silva, C. M., Gnanamanickam, E. P., & Atkinson, C. (2014). High spatial range velocity measurements in a high Reynolds number turbulent boundary layer. *Phys. Fluids*, *26*(025117). https://doi.org/10.1063/1.4866458

Dennis, D. J. C., & Nickels, T. B. (2011). Experimental measurement of large-scale three-dimensional structures in a turbulent boundary layer. Part 1. Vortex packets. *Journal of Fluid Mechanics*, *673*, 180-217. https://doi.org/10.1017/S0022112010006324

Deriaz, E., & Perrier, V. (2009). Direct Numerical Simulation of Turbulence Using Divergence-Free Wavelets. *Multiscale Modeling & Simulation*, *7*(3), 1101-1129. https://doi.org/10.1137/070701017

Ding, C. P., Peterson, B., Schmidt, M., Dreizler, A., & Böhm, B. (2019). Flame/flow dynamics at the piston surface of an IC engine measured by high-speed PLIF and PTV. *Proceedings of the Combustion Institute*, *37*(4), 4973-4981. https://doi.org/10.1016/J.PROCI.2018.06.215

Dormand, J. R., & Prince, P. J. (1980). A family of embedded Runge-Kutta formulae. *Journal of Computational and Applied Mathematics*, *6*(1), 19-26. https://doi.org/10.1016/0771-050X(80)90013-3

Dérian, P., Héas, P., Herzet, C., & Mémin, E. (2013). Wavelets and optical flow motion estimation. *Numerical Mathematics*, *6*(1), 1-23. https://doi.org/10.4208/nmtma.2013.mssvm07

Farge, M., Kevlahan, N., Perrier, V., & Goirand, É. (1996). Wavelets and turbulence. *Proceedings of the IEEE*, *84*(4), 639 - 669. https://doi.org/10.1109/5.488705

Gao, Q., Ortiz-Dueñas, C., & K. Longmire, E. (2013). Evolution of coherent structures in turbulent boundary layers based on moving tomographic PIV. *Experiments in Fluids*, *54*(12). https://doi.org/10.1007/s00348-013-1625-0

Gevelber, T. S., Schmidt, B. E., Mustafa, M. A., Shekhtman, D., & Parziale, N. J. (2022). Determining velocity from tagging velocimetry images using optical flow. *Experiments in Fluids*, *63*(6), 104-104. https://doi.org/10.1007/S00348-022-03448-Z

Graham, J., Kanov, K., Yang, X. I. A., Lee, M., Malaya, N., Lalescu, C. C., . . . Meneveau, C. (2016). A web services accessible database of turbulent channel flow and its use for testing a new integral wall model for LES. *Journal of Turbulence*, *17*(2), 181-215. https://doi.org/10.1080/14685248.2015.1088656

Heitz, D., Mémin, E., & Schnörr, C. (2010). Variational fluid flow measurements from image sequences: Synopsis and perspectives. *Experiments in Fluids*, *48*(3), 369-393. https://doi.org/10.1007/s00348-009-0778-3

Herpin, S., Stanislas, M., Foucaut, J. M., & Coudert, S. (2012). Influence of the Reynolds number on the vortical structures in the logarithmic region of turbulent boundary layers. *J. Fluid Mech*, *716*, 5-50. https://doi.org/10.1017/jfm.2012.491

Herpin, S., Wong, C. Y., Stanislas, M., & Soria, J. (2008). Stereoscopic PIV measurements of a turbulent boundary layer with a large spatial dynamic range. *Experiments in Fluids*, *45*(4), 745-763. https://doi.org/10.1007/s00348-008-0533-1

Heás, P., Herzet, C., Mémin, E., Heitz, D., & Mininni, P. D. (2013). Bayesian estimation of turbulent motion. *IEEE Transactions on Pattern Analysis and Machine Intelligence*, *35*(6), 1343-1356. https://doi.org/10.1109/TPAMI.2012.232

Horn, B. K. P., & Schunck, B. G. (1981). Determining optical flow. *Artificial Intelligence*, *17*(1-3), 185-203. https://doi.org/10.1016/0004-3702(81)90024-2

Héas, P., Mémin, E., Heitz, D., & Mininni, P. D. (2012). Power laws and inverse motion modelling: Application to turbulence measurements from satellite images. *Tellus, Series A: Dynamic Meteorology and Oceanography*, *64*(1). https://doi.org/10.3402/tellusa.v64i0.10962

Jainski, C., Lu, L., Dreizler, A., & Sick, V. (2013). High-speed micro particle image velocimetry studies of boundary-layer flows in a direct-injection engine. *International Journal of Engine Research*, *14*(3), 247-259. https://doi.org/10.1177/1468087412455746

Jainski, C., Rißmann, M., Jakirlic, S., Böhm, B., & Dreizler, A. (2018). Quenching of Premixed Flames at Cold Walls: Effects on the Local Flow Field. *Flow, Turbulence and Combustion*, *100*(1), 177–196. https://doi.org/10.1007/s10494-017-9836-8

Kadri-Harouna, S., Dérian, P., Héas, P., & Mémin, E. (2013). Divergence-free wavelets and high order regularization. *International Journal of Computer Vision*, *103*(1), 80-99. https://doi.org/10.1007/s11263-012-0595-7

Kalmoun, E. M. (2018). An investigation of smooth TV-like regularization in the context of the optical flow problem. *Journal of Imaging*, *4*(2). https://doi.org/10.3390/jimaging4020031

Kapulla, R., Hoang, P., Szijarto, R., & Fokken, J. (2011). Parameter sensitivity of optical flow applied to PIV Images. *Proceedings of the Fachtagung "Lasermethoden in der Strömungsmesstechnik", Ilmenau, Germany*, 6-8.

Keane, R. D., Adrian, R. J., & Zhang, Y. (1995). Super-resolution particle imaging velocimetry. *Measurement Science and Technology*, *6*(6). https://doi.org/10.1088/0957-0233/6/6/013

Kosaka, H., Zentgraf, F., Scholtissek, A., Bischoff, L., Häber, T., Suntz, R., . . . Dreizler, A. (2018). Wall heat fluxes and CO formation/oxidation during laminar and turbulent side-wall quenching of methane and DME flames. *International Journal of Heat and Fluid Flow*, *70*, 181-192. https://doi.org/10.1016/j.ijheatfluidflow.2018.01.009

Kosaka, H., Zentgraf, F., Scholtissek, A., Hasse, C., & Dreizler, A. (2020). Effect of Flame-Wall Interaction on Local Heat Release of Methane and DME Combustion in a Side-Wall Quenching Geometry. *Flow, Turbulence and Combustion*, *104*(4), 1029-1046. https://doi.org/10.1007/s10494-019-00090-4

Kähler, C. J., Astarita, T., Vlachos, P. P., Sakakibara, J., Hain, R., Discetti, S., . . . Cierpka, C. (2016). Main results of the 4th International PIV Challenge. *Experiments in Fluids*, *57*(6), 1-71. https://doi.org/10.1007/s00348-016-2173-1

Kähler, C. J., Scharnowski, S., & Cierpka, C. (2012a). On the resolution limit of digital particle image velocimetry. *Experiments in Fluids*, *52*(6), 1629–1639. https://doi.org/10.1007/s00348-012-1280-x

Kähler, C. J., Scharnowski, S., & Cierpka, C. (2012b). On the uncertainty of digital PIV and PTV near walls. *Experiments in Fluids*, *52*(6), 1641-1656. https://doi.org/10.1007/s00348-012-1307-3

Lehew, J. A., Guala, M., & McKeon, B. J. (2013). Time-resolved measurements of coherent structures in the turbulent boundary layer. *Experiments in Fluids*, *54*(4), 1-16. https://doi.org/10.1007/s00348-013-1508-4

Li, Y., Perlman, E., Wan, M., Yang, Y., Meneveau, C., Burns, R., . . . Eyink, G. (2008). A public turbulence database cluster and applications to study Lagrangian evolution of velocity increments in turbulence. *Journal of Turbulence*, *9*. https://doi.org/10.1080/14685240802376389

Liu, T., & Shen, L. (2008). Fluid flow and optical flow. *Journal of Fluid Mechanics*, *614*, 253-291. https://doi.org/10.1017/S0022112008003273

Lu, J., Yang, H., Zhang, Q., & Yin, Z. (2021). An accurate optical flow estimation of PIV using fluid velocity decomposition. *Experiments in Fluids*, *62*(4), 1-16. https://doi.org/10.1007/s00348-021-03176-w

Mallat, S. (2009). *A Wavelet Tour of Signal Processing*. https://doi.org/10.1016/B978-0-12-374370-1.X0001-8

McCane, B., Novins, K., Crannitch, D., & Galvin, B. (2001). On benchmarking optical flow. *Computer Vision and Image Understanding*, *84*(1), 126-143. https://doi.org/10.1006/cviu.2001.0930

Ouyang, Z., Yang, H., Huang, Y., Zhang, Q., & Yin, Z. (2021). A circulant-matrix-based hybrid optical flow method for PIV measurement with large displacement. *Experiments in Fluids*, *62*(11), 1-18. https://doi.org/10.1007/S00348-021-03317-1

Pope, S. B. (2002). Turbulent Flows (Cambridge University Press, UK).

Prandtl, L. (1904). Über Flüssigkeitsbewegung bei sehr kleiner Reibung ("Motion of Fluids with Very Little Viscosity"). *Internationalen Mathematiker-Kongresses*, 484-491.

Raffel, M., Willert, C. E., Scarano, F., Kähler, C. J., Wereley, S. T., & Kompenhans, J. (2018). *Particle Image Velocimetry: A Practical Guide*.

Renaud, A., Ding, C. P., Jakirlic, S., Dreizler, A., & Böhm, B. (2018). Experimental characterization of the velocity boundary layer in a motored IC engine. *International Journal of Heat and Fluid Flow*, *71*, 366-377. https://doi.org/10.1016/J.IJHEATFLUIDFLOW.2018.04.014

Robinson, S. K. (1991). Coherent motions in the turbulent boundary layer. *Annual Review of Fluid Mechanics*, *23*(1), 601-639. https://doi.org/10.1146/ANNUREV.FL.23.010191.003125

Ruhnau, P., & Schnörr, C. (2006). Optical Stokes flow estimation: an imaging-based control approach. *Experiments in Fluids*, *42*(1), 61-78. https://doi.org/10.1007/s00348-006-0220-z

Ruhnau, P., Stahl, A., & Schnörr, C. (2007). Variational estimation of experimental fluid flows with physics-based spatio-temporal regularization. *Measurement Science and Technology*, *18*(3). https://doi.org/10.1088/0957-0233/18/3/027

Scarano, F., & Riethmuller, M. L. (2000). Advances in iterative multigrid PIV image processing. *Experiments in Fluids*, *29*(1), 51-60.

Schmidt, B. E., Skiba, A. W., Hammack, S. D., Carter, C. D., & Sutton, J. A. (2021). High-resolution velocity measurements in turbulent premixed flames using wavelet-based optical flow velocimetry (wOFV). Proceedings of the Combustion Institute,

Schmidt, B. E., & Sutton, J. A. (2019). High-resolution velocimetry from tracer particle fields using a wavelet-based optical flow method. *Experiments in Fluids*, *60*(3). https://doi.org/10.1007/s00348-019-2685-6

Schmidt, B. E., & Sutton, J. A. (2020). Improvements in the accuracy of wavelet-based optical flow velocimetry (wOFV) using an efficient and physically based implementation of velocity regularization. *Experiments in Fluids*, *61*(2). https://doi.org/10.1007/s00348-019-2869-0

Schmidt, B. E., & Sutton, J. A. (2021). A physical interpretation of regularization for optical flow methods in fluids. *Experiments in Fluids*, *62*(2). https://doi.org/10.1007/s00348-021-03147-1

Schmidt, B. E., & Woike, M. R. (2021). Wavelet-based optical flow analysis for background-oriented schlieren image processing. *AIAA Journal*, *59*(8). https://doi.org/10.2514/1.J060218

Schmidt, M., Ding, C.-P., Peterson, B., Dreizler, A., & Böhm, B. (2021). Near-Wall Flame and Flow Measurements in an Optically Accessible SI Engine. *106*, 597-611. https://doi.org/10.1007/s10494-020-00147-9

Schröder, A., Geisler, R., Staack, K., Elsinga, G. E., Scarano, F., Wieneke, B., . . . Westerweel, J. (2011). Eulerian and Lagrangian views of a turbulent boundary layer flow using time-resolved tomographic PIV. *Experiments in Fluids*, *50*(50), 1071-1091. https://doi.org/10.1007/s00348-010-1014-x

Spalart, P. R. (1988). Direct simulation of a turbulent boundary layer up to R, = 1410. *J. Fluid Mech*, *187*, 61-98. https://doi.org/10.1017/S0022112088000345

Stanislas, M., Okamoto, K., Kähler, C. J., & Westerweel, J. (2005). Main results of the Second International PIV Challenge. Experiments in Fluids,

Stark, M. (2013). *Optical Flow PIV: Improving the Accuracy and Applicability of Particle Image Velocimetry*. ETH, Department of Mechanical and Process Engineering.

Stitou, A., & Riethmuller, M. L. (2001). Extension of PIV to super resolution using PTV. *Meas. Sci. Technol*, *12*, 1398-1403.

Suter, D. (1994). Motion estimation and vector splines. Proceedings of the IEEE Computer Society Conference on Computer Vision and Pattern Recognition,

Weickert, J., & Schnörr, C. (2001). A theoretical framework for convex regularizers in PDE-based computation of image motion. *International Journal of Computer Vision*, *45*(3), 245-264. https://doi.org/10.1023/A:1013614317973

Willert, C. E. (2015). High-speed particle image velocimetry for the efficient measurement of turbulence statistics. *Exp Fluids*, *56*(1), 17-17. https://doi.org/10.1007/s00348-014-1892-4

Winger, L. L., & Venetsanopoulos, A. N. (2001). Biorthogonal nearly coiflet wavelets for image compression. *Signal Processing: Image Communication*, *16*(9), 859-869. https://doi.org/10.1016/S0923-5965(00)00047-3

Wu, Y. T., Kanade, T., Li, C. C., & Cohn, J. (2000). Image registration using wavelet-based motion model. *International Journal of Computer Vision*, *38*(2), 129-152. https://doi.org/10.1023/A:1008101718719

Yuan, J., Schnörr, C., & Mémin, E. (2007). Discrete orthogonal decomposition and variational fluid flow estimation. *Journal of Mathematical Imaging and Vision*, *28*(1), 67-80. https://doi.org/10.1007/s10851-007-0014-9

Zach, C., Pock, T., & Bischof, H. (2007). A Duality Based Approach for Realtime TV-{\itshape L}<Superscript>1</Superscript> Optical Flow. Pattern Recognition,

Zentgraf, F. (2022). *Investigation of Reaction and Transport Phenomena during Flame-Wall Interaction Using Laser Diagnostics.* Dissertation, Technical University of Darmstadt.

Zentgraf, F., Johe, P., Cutler, A. D., Barlow, R. S., Böhm, B., & Dreizler, A. (2021). Classification of flame prehistory and quenching topology in a side-wall quenching burner at low-intensity turbulence by correlating transport effects with CO2, CO and temperature. *Combustion and Flame*(239). https://doi.org/10.1016/j.combustflame.2021.111681

Zentgraf, F., Johe, P., Steinhausen, M., Hasse, C., Greifenstein, M., Cutler, A. D., . . . Dreizler, A. (2022). Detailed assessment of the thermochemistry in a side-wall quenching burner by simultaneous quantitative measurement of CO2, CO and temperature using laser diagnostics. *Combustion and Flame*, *235*. https://doi.org/10.1016/j.combustflame.2021.111707

Zhang, X. Y., Wang, L. M., Liu, B., Luo, Y., & Han, X. C. (2020). Hybrid Adaptive Wavelet-Based Optical Flow Algorithm for Background Oriented Schlieren (BOS) Experiments. *Mathematical Problems in Engineering*, *2020*. https://doi.org/10.1155/2020/5138153